\documentclass[prb,aps,showpacs,twocolumn,preprintnumbers,
amsmath,amssymb,superscriptaddress,longbibliography]{revtex4-2}

\usepackage[english]{babel}
\usepackage{amsmath,amssymb,amsfonts}
\usepackage{graphicx}
\usepackage[colorlinks=True,linkcolor=red,citecolor=blue,urlcolor=blue]{hyperref}

\usepackage{booktabs}
\usepackage{longtable}

\usepackage[dvipsnames]{xcolor}
\usepackage{verbatim}
\usepackage{braket}
\usepackage{bm}
\usepackage{bbm}
\usepackage{braket}
\usepackage{float}
\usepackage{multirow}
\usepackage[normalem]{ulem}
\usepackage{array}
\usepackage{makecell}
\usepackage{subfigure}
\usepackage{float}
\usepackage{xcolor}
\newcolumntype{C}{>{$}c<{$}}

\newcommand{\abs}[1]{\left\lvert #1 \right\rvert}

\allowdisplaybreaks
\DeclareMathAlphabet{\zc}{OT1}{pzc}{m}{it}

\definecolor{jade}{rgb}{0.0, 0.66, 0.42}
\def\changes{\textcolor{black}}

\begin{document}

\title{Beyond the non-Hermitian skin effect: scaling-controlled topology from Exceptional-Bound Bands}

\author{Mengjie Yang}
\email{mengjie.yang@u.nus.edu}
\affiliation{Department of Physics, National University of Singapore, Singapore 117551, Singapore}

\author{Ching Hua Lee}
\email{phylch@nus.edu.sg}
\affiliation{Department of Physics, National University of Singapore, Singapore 117551, Singapore}

\date{\today}

\begin{abstract}
We establish a novel mechanism for topological transitions in non-Hermitian systems that are controlled by the system size. Based on a new paradigm known as exceptional-bound (EB) band engineering, its mechanism hinges on the unique critical scaling behavior near an exceptional point, totally unrelated to the well-known non-Hermitian skin effect. Through a series of ansatz models, we analytically derive and numerically demonstrate how topological transitions depend on the system size with increasingly sophisticated topological phase boundaries. Our approach can be generically applied to design scaling-dependent bands in multi-dimensional lattices, gapped or gapless, challenging established critical and entanglement behavior. It can be experimentally demonstrated in any non-Hermitian platform with versatile couplings or multi-orbital unit cells, such as photonic crystals, as well as classical and quantum circuits. The identification of this new EB band mechanism provides new design principles for engineering band structures through scaling-dependent phenomena unique to non-Hermitian systems.
\end{abstract}

\maketitle
\newpage

\section{Introduction}

Traditionally, topological phase transitions have been understood as phenomena emerging in the thermodynamic limit. However, in non-Hermitian systems, it has become clear that the same system at different finite sizes can possess distinct topological characters, hosting robust edge states only for sufficiently small or sufficiently large system sizes~\cite{li2020critical,liu2020helical,yokomizo2021scaling,arouca2020unconventional,rafi2022unconventional,kawabata2023entanglement,qin2023universal,yang2024percolation}. This observation carries direct experimental implications, as realistic setups are inevitably finite and bounded.

In studies so far, the mechanism underlying such scale-dependent topological transitions has been competition between different localization channels induced by the non-Hermitian skin effect~\cite{lee2016anomalous,yao2018edge,lee2019anatomy,lee2019hybrid,longhi2019topological,okuma2020topological,yang2025non,yang2025reversing}. With skin state accumulation growing exponentially with distance, larger system sizes would unleash new state tunneling routes, qualitatively changing the connectivity of the system and its topological band structure~\cite{li2020critical,yang2024percolation}.
Such non-Hermitian scaling behavior has attracted much attention by challenging traditional notions of criticality and quantum entanglement properties~\cite{longhi2019topological,luo2021transfer,kawabata2023entanglement,liu2024non}.

In this work, however, we show that anomalous non-Hermitian scaling can occur \emph{without} any skin effect at all.
Departing from existing paradigms~\cite{li2020critical,liu2020helical,arouca2020unconventional,liu2020helical,guo2021exact,yokomizo2021scaling,rafi2022system,rafi2022critical,qin2023universal,yang2024percolation,yang2025beyond}, we propose a completely different mechanism for scale-induced topological transitions.
Our approach relies on unique renormalization properties near an exceptional point (EP) due to eigenspace defectiveness.
\changes{While EPs are already known for their extremely sensitive band dispersions and enhanced response in sensing-oriented settings}~\cite{dembowski2001experimental,hodaei2017enhanced,chen2017exceptional,ozdemir2019parity,miri2019exceptional,heiss2012physics,shi2016accessing,meng2024exceptional,li2023exceptional,goldzak2018light,heiss2004exceptional,zhang2024true} and unconventional entanglement properties~\cite{gopalakrishnan2021entanglement,fossati2023symmetry,lee2022exceptional,chen2022quantum,turkeshi2023entanglement,li2024emergent,xue2024topologically,orito2022unusual}, what has not been previously harnessed is the non-locality that emerges from projecting onto their defective eigenspaces.

To employ this non-locality for realizing novel scaling transitions, we specifically \emph{dimensionally extend} these eigenspaces to form so-called \emph{exceptional-bound (EB) bands}. These are entire energy bands protected by the presence of momentum-space singularities. As we shall show, EB bands possess unique algebraically decaying profiles that scale strongly with system size, harboring the propensity for scale-sensitive band engineering.
\changes{Starting from a parent exceptional point, we demonstrate how we can build effective models with precisely scale-tunable couplings.
Unlike long-range coupling \emph{design} approaches that tune hopping profiles link-by-link at fixed size~\cite{chen2024ultra,guo2024scale},
our long-range couplings are uniquely generated once the EP projector kernel and truncation rule are specified, with $L_y$ serving as the primary control knob.} This enables the same system to be topological at some system sizes but non-topological at other sizes, i.e., exhibiting scaling-controlled topological phase transitions. We develop a formalism for engineering such unprecedented scale-controlled transitions, fundamentally expanding the potential of non-Hermitian mechanisms in non-locally controlling entire band structures, beyond the reach of existing sensing application proposals.

\section{Results}
\subsection{Unconventional scaling behavior from exceptional point defectiveness}

To explain our new mechanism for scaling-controlled topology, we first show how projecting onto a geometrically defective exceptional point (EP) eigenspace can give rise to strongly scale-dependent eigenstates.
Consider a minimal EP model
\begin{equation}
\mathcal{H}(k)=\left(
\begin{array}{cc}
  0 & a_0+h(k) \\
  h(k) & 0
\end{array}\right), \label{eq:H2k}
\end{equation}
where $h(k)\sim k^{2B}\rightarrow 0$ as $k\rightarrow 0$, with $B>0$ a positive integer representing the order of the EP. When $a_0 \neq 0$, the two eigenvectors of $\mathcal{H}(k)$ coalesce as $k\rightarrow 0$, leading to a geometrically defective eigenspace. This defectiveness is most saliently captured by the biorthogonal projector onto the lower band, as given by~\cite{okuma2021quantum,orito2022unusual,kawabata2023entanglement,li2024emergent}
\begin{equation}
P(k)=\sum_{n\in occ.}\ket{\psi_{n}(k)}\bra{\tilde{\psi}_{n}(k)}=\frac{1}{2}\left(
\begin{array}{cc}
  1 & -U(k) \\
  -D(k) & 1
\end{array}\right), \label{eq:P2k}
\end{equation}
\begin{eqnarray}
U(k)&=\sqrt{\frac{a_0+h(k)}{h(k)}},\\
D(k)&=\sqrt{\frac{h(k)}{a_0+h(k)}}.
\label{UDk}
\end{eqnarray}
Importantly, the off-diagonal correlator $U(k)=D(k)^{-1}\sim a_0k^{-B}$ diverges as $k\rightarrow 0$, since the projector should be ill-defined when the eigenvectors coalesce. Note that in order to satisfy $P^2=P$ in a non-Hermitian setting, this projector $P$ must be defined with respect to the biorthogonal basis which is given by $\mathcal{H}\ket{\psi_{n}}=\mathcal{E}\ket{\psi_{n}}$ and $\mathcal{H}^\dagger\ket{\tilde{\psi}_{n}}=\mathcal{E}^*\ket{\tilde{\psi}_{n}}$, with $\langle\tilde{\psi}_{n}\ket{\psi_{n'}}=\delta_{nn'}$.

Such divergences in the projector $P(k\rightarrow 0)$ lead to highly non-local correlations in real space with dramatic scaling implications. Specifically, the real space propagators (2-point correlators) across $y$ unit cells within the lower band are given by $U_y=\frac1{L_y}\sum_{k}U(k)e^{iky}$ and $D_y=\frac1{L_y}\sum_{k}D(k)e^{iky}$, which are the Fourier transforms~\cite{lee2015free,carlstrom2020correlations} of $U(k)$ and $D(k)$.
Using the lattice regularization
\begin{equation}
h(k)=\frac{1}{2}(2(1-\cos k))^B
\end{equation}
for definiteness, we have
\begin{equation}
\begin{aligned}
U_y& \approx\frac{\sqrt{a_0}}{2^{B-2}} \int_{\pi / 2 L_y}^{\pi / 2}d k^{\prime}\frac{\cos 2k' y}{\sin ^{B} k^{\prime}},\\
D_y&\approx\frac{2^{B+2}}{\sqrt{a_0}} \int_{\pi / 2L_y}^{\pi/2}d k' \sin^B k' \cos 2k' y,
\end{aligned}\label{eq:Uy}
\end{equation}
where the lower integration limit $\pi/2L_y$ scales inversely proportional to system size $L_y$.
As we later show [Eq.~\eqref{eq:Uy_B123}, due to the divergence of $U(k\rightarrow 0)$, $U_y$ diverges with $L_y$, behaving like slowly-decaying long-range propagators. By contrast, $D(k)=U(k)^{-1}\rightarrow 0$ as $k\rightarrow 0$ is not divergent, and its Fourier coefficients $D_y$ decay rapidly, as elaborated in Methods Sect.~\ref{method:EBNN}.

\begin{figure}
\centering
\includegraphics[width=0.48\textwidth]{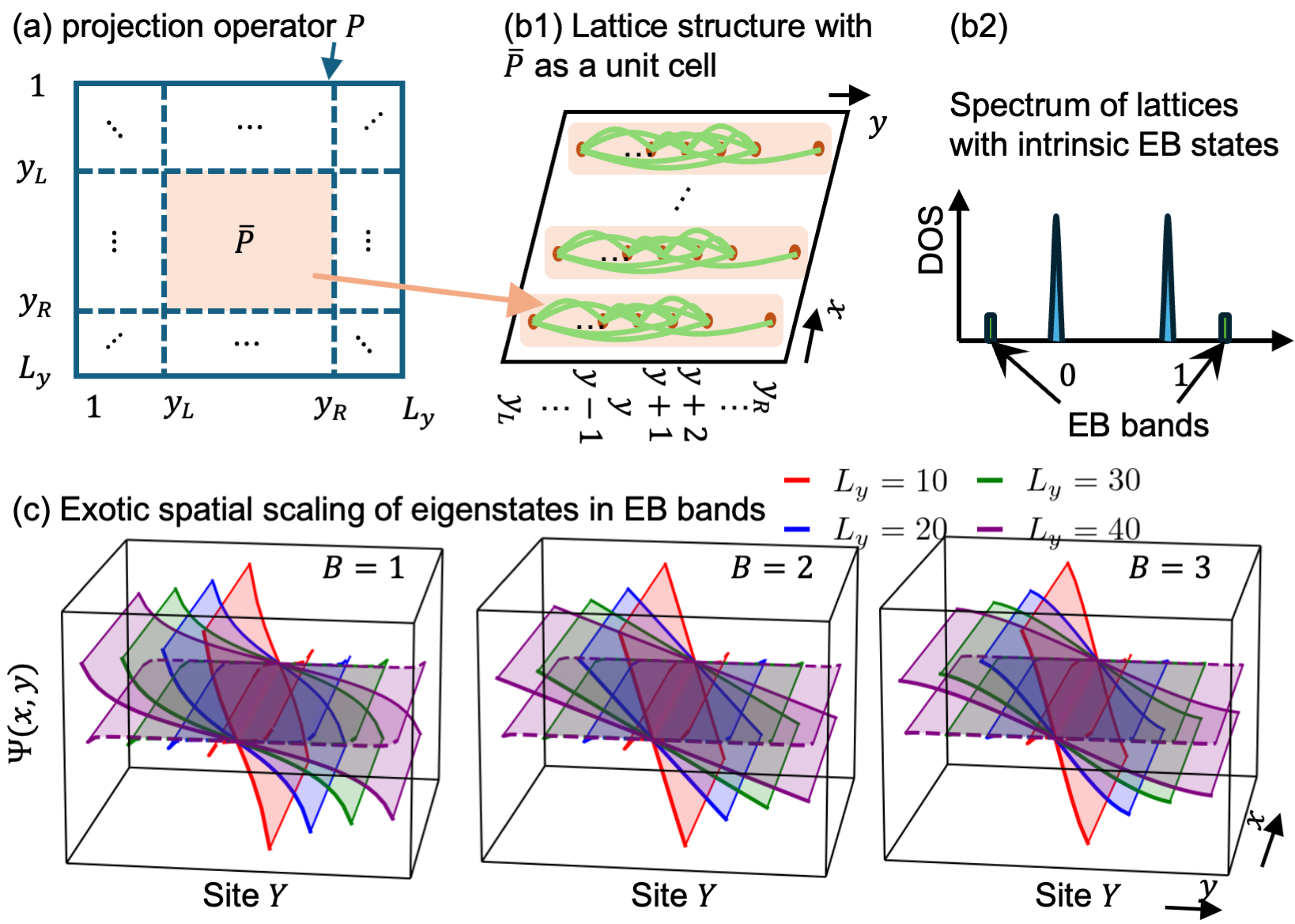}
\caption{Construction of exceptional-bound (EB) band lattice models from parent exceptional points (EPs). (a) From the band projector $P$ (exterior blue square) of a parent EP, one can construct a $\bar P$ operator [Eq.~\eqref{eq:barP}] by truncating the degrees of freedom outside of a chosen interval $y\in [y_L,y_R]$ (orange square).
  (b1) To craft lattice models featuring intrinsic EB bands, one defines $y_R-y_L+1$-atom unit cells whose connectivity structure (green) is given by the $\bar P$ matrix elements. Their long-ranged nature~\cite{lee2022exceptional,zou2024experimental}, inherited from the defectiveness of the parent EP, gives rise to unconventional scaling properties. (b2) Also due to this EP defectiveness, the spectrum of the lattice in (b1) features a pair of robust eigenvalues beyond the usual bounds of 0 and 1, which we call EB bands.
(c) Illustrative EB band basis eigenstates corresponding to different parent EP orders $B=1,2,3$, with solid/dashed surfaces representing state amplitudes on odd/even sites, computed with $y_L=1,y_R=L_y-1$. Pronounced scaling behavior arises from the very different basis eigenstates profiles at different system sizes $L_y=10,20,30,40$ (red, blue, green, purple, respectively).}
\label{fig:barPunitcell}
\end{figure}

In a bounded physical setting, the divergently long-ranged nature of the propagators $U_y$
can profoundly change part of the spectrum (band structure) in a unique manner, as we motivate below.
Restricted to a bounded subsystem $\Gamma=[y_L,y_R]$ via a real space projection $R_\Gamma$ [Fig.~\ref{fig:barPunitcell}(a)], the biorthogonal projector becomes
\begin{equation}
\begin{aligned}
  \bar P&=\sum_{y,y'\in\Gamma}\left[L_y^{-1}\sum_k e^{i k(y-y')} P(k)\right]|y'\rangle  \langle y| \\
  &=\frac1{2}\sum_{y,y'\in\Gamma}\left(
    \begin{array}{cc}
      \delta_{yy'} & -U_{y-y'} \\
      -D_{y-y'}  & \delta_{yy'}
  \end{array}\right)|y'\rangle  \langle y|\\
  &=\frac1{2}\left(
    \begin{array}{cc}
      R_\Gamma & -R_\Gamma UR_\Gamma \\
      -R_\Gamma DR_\Gamma  & R_\Gamma
  \end{array}\right).
\end{aligned}
\label{eq:barP}
\end{equation}
Interestingly, even though $P$ is a projector, i.e., $P^2-P=0$, $\bar P$ can deviate significantly from being a projector. To see that, note that $\bar P^2- \bar P=$
\begin{equation}
\begin{aligned}
  &\frac1{4}\left(
    \begin{array}{cc}
      R_\Gamma+R_\Gamma U R_\Gamma D R_\Gamma & -2R_\Gamma UR_\Gamma \\
      -2 DR_\Gamma DR_\Gamma & R_\Gamma+R_\Gamma D R_\Gamma U R_\Gamma
  \end{array}\right)-\bar P\\
  &=\frac1{2}\left(
    \begin{array}{cc}
      R_\Gamma U R_\Gamma D R_\Gamma -R_\Gamma &  0\\
      0  & R_\Gamma D R_\Gamma U R_\Gamma -R_\Gamma
  \end{array}\right)\\
  &= -\frac1{2}R_\Gamma\left(
    \begin{array}{cc}
      UR^c_\Gamma U^{-1} &  0\\
      0 & U^{-1}R^c_\Gamma U
  \end{array}\right)R_\Gamma\\
  &= -\frac1{2}\sum_{y,y'\in\Gamma}|y'\rangle  \langle y|  \left[\sum_{y''\notin\Gamma}\left(
      \begin{array}{cc}
        U_{y''-y}D_{y'-y''}& 0\\
        0 & D_{y''-y}U_{y'-y''}
  \end{array}\right)\right]
\end{aligned}
\label{eq:barP2}
\end{equation}
where $R^c_\Gamma=\mathbb{I}-R_\Gamma$, and we have used $R^2_\Gamma=R_\Gamma$ and $U^{-1}=D$.
Evidently, $\bar P^2= \bar P$ is violated to a significant extent due to divergently large propagators $U_{y''-y}$ between $y\in \Gamma$ and $y''\notin \Gamma$, which in turn result from the geometrically defective EP.

The key consequence of the failure of $\bar P$ as a projector (i.e. $\bar P^2\neq \bar P$) is that it hosts a pair of robust isolated ``exceptional-bound (EB)'' eigenvalues $\bar p$, $1-\bar p$ that significantly deviates from $0$ and $1$ [Fig.~\ref{fig:barPunitcell}(b)]. They differ from usual projector eigenstates with eigenvalues $0$ and $1$, as well as in-gap topological states that can lie within [0,1]~\cite{thomale2010entanglement,fidkowski2010entanglement,hughes2011inversion,qi2012general,hermanns2014entanglement}.
In the following, we demonstrate how such EB states can form the building block of novel 2D band structures with intrinsic scaling-dependent properties.

\subsection{Exceptional-bound band engineering: Approach and Examples}

Central to our approach is the construction of very robust bands — we term these ``exceptional-bound (EB) bands'' — intrinsically equipped with unconventional scaling properties. Built from robustly gapped EB states, they are very different from topological or non-Hermitian skin states, which are typically exponentially localized. Instead, EB bands would exhibit unique spatial profiles that can be harnessed for scale-dependent band engineering.

Starting from a 2-band ansatz, i.e., Eq.~\eqref{eq:P2k} parametrized by integer order $B$, we can define an EB band lattice [Fig.~\ref{fig:barPunitcell}(b)] with each unit ``supercell'' formed from its corresponding $\bar{P}$ [Eq.~\eqref{eq:barP}]. Since $\bar P$ is a $[2(y_R-y_L+1)] \times [2(y_R-y_L+1)]$ matrix, each matrix element of $\bar P$ corresponds to the coupling strength (green) between the $2(y_R-y_L+1)$ different atoms within the unit cell. This construction sets the stage for EB band engineering, starting from a foundational lattice with pre-built EB states in each unit cell. While the intra-unit-cell couplings are inevitably non-local, the total number of non-local couplings can be truncated down to reasonable numbers in the ballpark of $10-10^2$, since EB eigenvalues persist robustly
even as the cut region $[y_L, y_R]$ is shrunk to less than $\mathcal{O}(10)$.

Shown in Fig.~\ref{fig:barPunitcell}(c) are the profiles of illustrative EB bands $\Psi(x,y)$ in a lattice of $\bar P$ unit cells. The EB bands satisfy the eigenequation $\bar P\Psi=\bar p \Psi$, where $x$ labels the $\bar P$ unit cell in an extended $x$ dimension, and $y$ indexes the ``internal'' space within each unit cell. As will be derived in Sect.~\ref{sect:EBscaling}, the EB state profiles differ qualitatively for different parent exceptional point orders $B=1,2$ and $3$ [Eq.~\eqref{eq:H2k}], and importantly possess dramatic scaling behavior with $L_y=10,20,30,40$ (red, blue, green, purple, respectively). In the following, we shall explicitly demonstrate how this pronounced and unique scaling behavior can give rise to EB band structures that behavior qualitatively differently at different $L_y$.

\subsubsection{Warm-up: scale-independent EB topological bands }

\begin{figure}
\centering
\includegraphics[width=0.48\textwidth]{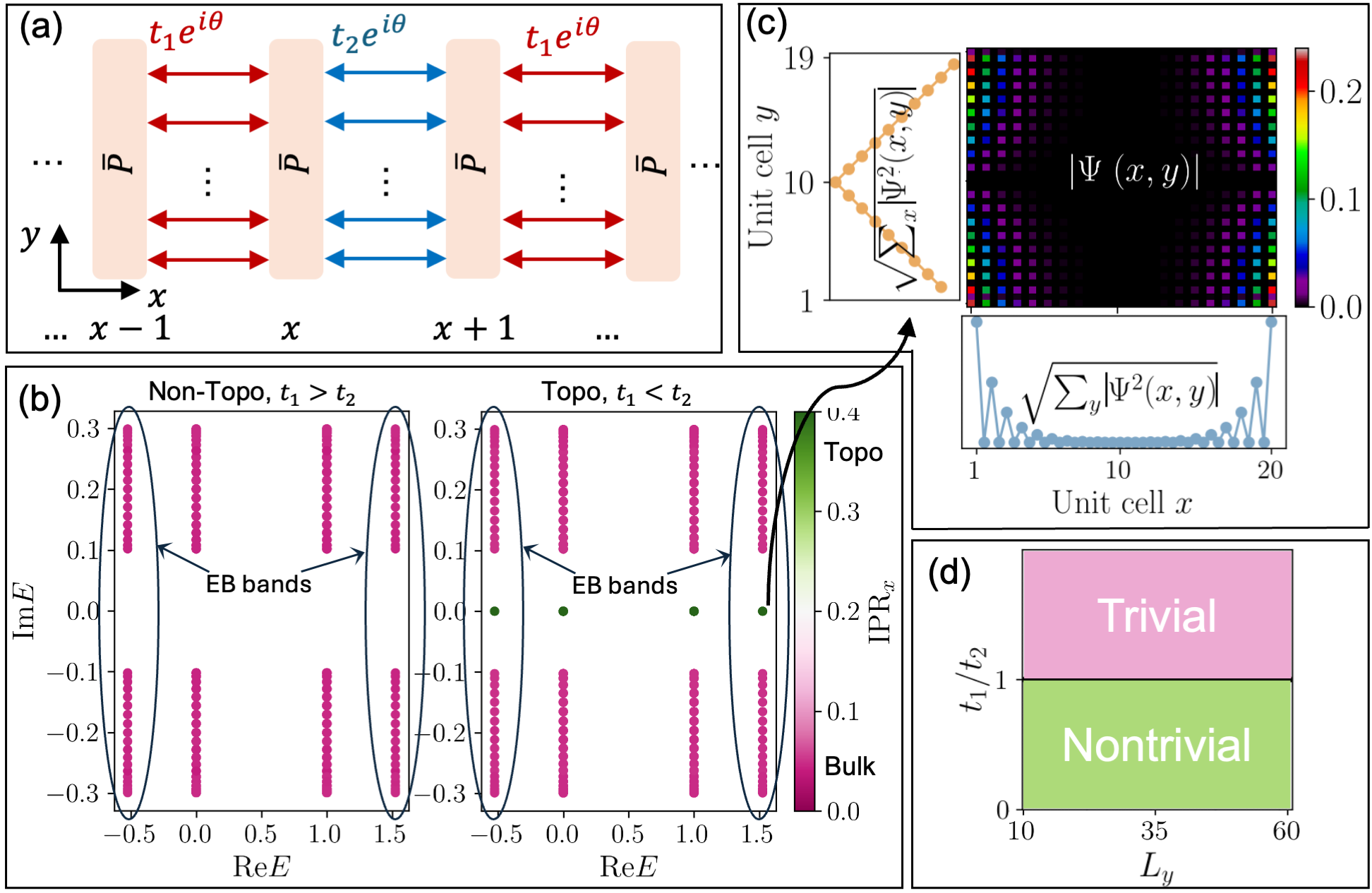}
\caption{Topological characterization of the NN EB-SSH warm-up model in Eq.~\eqref{eq:ham_SSH_NN}, as a prelude to the scale-dependent transition scenario in Fig.~\ref{fig:phasediagram}. (a) Lattice structure, showing alternating hopping amplitudes $t_1$ and $t_2$ between $\bar P$ unit cell building blocks. (b)
  Band structure obtained from $H_{\text{NN}}^{\text{(EBB)}}\Psi_i(x,y)=E_i\Psi_i(x,y)$, with angle $\theta=\pi/2$ between lattice (x) and EB (y) degrees of freedom. Mid-gap topological states (green), which appear only for $t_1<t_2$, can be distinguished from the bulk states (pink) through IPR$_x$ [Eq.~\eqref{eq:IPR_x}]. EB bands (circled) do not play special roles in the topological transition within this warm-up scenario.
  (c) Spatial state distribution $|\Psi(x,y)|$ for a topological EB eigenstate at $E\approx 1.5$, which simultaneously incorporates topological (blue) and EB (orange) localization with dissimilar exponential $(t_1/t_2)^x$ and linear $y$ profiles, respectively.
  (d) In this warm-up model with trivially horizontal NN couplings, the topological phase boundary remains at the conventional $t_1=t_2$ line, with no scaling ($L_y$)-dependence.
Here, we used $B=2, a_0=10, y_R=L_y-1=19$. }
\label{fig:SSH_NN}
\end{figure}

\begin{figure*}
\centering
\includegraphics[width=0.98\textwidth]{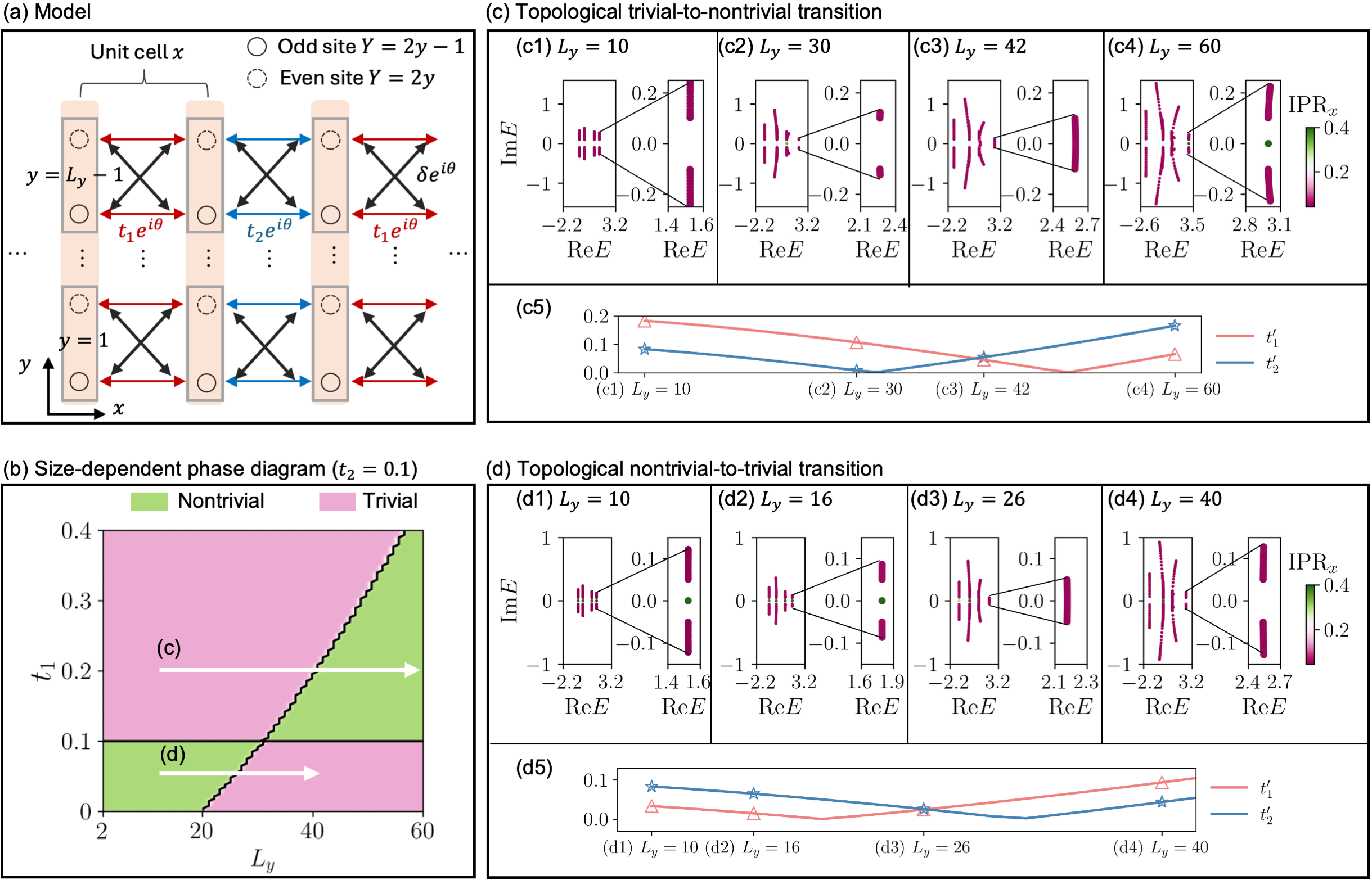}
\caption{The system size $L_y$ as a decisive control knob for topological phase transitions within our extended EB band model [Eq.~\eqref{eq:ham_SSH_NNN}].
  (a) Schematic of its lattice, whose sensitivity to length $L_y$ arises from the criss-crossing NNN couplings $\delta$ between unit cells $\bar P$.
  (b) Topological phase diagram as a function of hopping strength $t_1$ and $L_y=y_R+1$, for fixed $t_2$. Due to the unconventional scaling of EB bands, a strongly $L_y$-dependent phase boundary curve appears, as determined by the ratio $|t_1'/t_2'|$ [Eq.~\eqref{t1pt2p}]. As such, increasingly $L_y$ can enigmatically drive the topological transition in either direction, depending on $t_1/t_2$. Consequently, at a fixed $L_y$, there exists non-monotonic topologically trivial$\rightarrow$non-trivial$\rightarrow$ trivial transitions (or vice-versa).
(c,d) Scaling-induced topological transitions in either direction: (c) trivial$\rightarrow$nontrivial and (d) nontrivial$\rightarrow$trivial, as indicated by the white arrows in (b). These transitions can be accurately explained as crossings of the effective $\delta$-shifted EB couplings $t_1'$ and $t_2'$ [Eq.~\eqref{eq:effectivet1t2}], as shown in panels (c5) and (d5).}
\label{fig:phasediagram}
\end{figure*}

As a warm-up, we first show how we can construct the simplest possible topological EB band by trivially coupling $\bar P$ unit cells with period-2 modulation. This is done by replacing the atoms in the nearest-neighbor (NN) SSH model with $\bar P$ unit cells~[Fig.~\ref{fig:SSH_NN}(a)]. These $\bar{P}$ building blocks are linked exclusively through odd ($t_1$) and even bonds ($t_2$) of differing strengths (for ease of notation, we henceforth set $y_L=1$):
\begin{equation}
\begin{aligned}
  H^{(\text{EBB})}_{\text{NN}}&(k_x) =\mathbb{I}_{2} \otimes \bar P \\
  &+e^{i\theta}\left(
    \begin{array}{cc}
      0 & t_1+t_2 e^{ik_x}\\
      t_1+t_2 e^{-ik_x} & 0
  \end{array}\right) \otimes \mathbb{I}_{2y_R}.
\end{aligned}\label{eq:ham_SSH_NN}
\end{equation}
Fig.~\ref{fig:SSH_NN}(b) showcases the EB band spectra for topologically trivial ($t_1>t_2$) and nontrivial ($t_1<t_2$) cases under $x$-open boundary conditions (OBCs), colored by the $x$-direction localization of the corresponding eigenstate as defined by the inverse participation ratio~\cite{kramer1993localization,evers2008anderson}
\begin{equation}
\operatorname{IPR}_x=\sum_x\left(\sum_y|\Psi(x, y)|^2\right)^2
\label{eq:IPR_x}
\end{equation}
For our choice of $\theta=\pi/2$, the bulk bands (pink) extend along the $\text{Im}\,E$ direction, accompanied by midgap localized topological modes (green) with higher IPR$_x$ for $t_1<t_2$. These bands inherit their $\text{Re}\,E$ values from the $\bar P$ spectrum [Fig.~\ref{fig:barPunitcell}(b)]: the bands at $\text{Re}\,E=0$ or $1$ are extensively ($y_R-y_L$ -- fold) degenerate, while the two isolated bands at  $\text{Re}\,E\approx -0.5$ and $1.5$ are doubly degenerate EB bands.
If necessary, $\theta$ can be tuned to obtain band structures of desired complex dispersions [see Methods Sect. A]

Even though this is just a trivial warm-up example, its relevance for scaling-dependent band engineering can already be seen by examining the spatial eigenstate profile in both the extended ($x$) and ``internal'' ($y$) directions. As shown in Fig.~\ref{fig:SSH_NN}(c) for an illustrative topological EB band with $E\approx 1.5$, the $y$-averaged root mean square (rms) state amplitude $\sqrt{\sum_{y} \Psi^2_i(x,y)}$ decays exponentially as $(t_1/t_2)^x$ and $(t_1/t_2)^{L_x-x}$ (blue) as expected for a SSH topological state. However, it also exhibits the characteristically linear EB-profile (shown here for $B=2$) in the $y$-direction, as indicated in the $x$-averaged rms plot (orange).

For this warm-up example, the NN couplings that link different $\bar P$ unit cells only connect atoms with the same $y$ (as in the $\mathbb{I}_{2y_R}$ term in Eq.~\eqref{eq:ham_SSH_NN}). Hence they couple neighboring EB states that scale identically, and cannot possibly capture the non-uniform scaling of the EB basis with $L_y=y_R+1$. This is shown in Fig.~\ref{fig:SSH_NN}(d), which confirms that the topological phase boundary is unaffected by $L_y$. Starting from the next example, we shall show how non-trivial $L_y$-scaling dependencies can be generically introduced by considering further couplings.

\subsubsection{Scale-dependent topological transitions from EB band engineering}

Having discussed the warm-up example with no scaling dependencies, we next present the simplest manner in which the EB bands exhibit non-trivial $L_y$ dependence. This involves introducing next-nearest-neighbor (NNN) couplings $\delta$ to $H^{(\text{EBB})}_{\text{NN}}$ [Eq.~\eqref{eq:ham_SSH_NN}], as illustrated in Fig.~\ref{fig:phasediagram}(a):
\begin{equation}
H^{(\text{EBB})}_{\text{NNN}}(k_x, \delta) = H^{(\text{EBB})}_{\text{NN}}+\delta\sigma_x\otimes \left(T_{-1}+T_{+1}\right),  \label{eq:ham_SSH_NNN}
\end{equation}
where $\delta\sigma_x\otimes \left(T_{-1}+T_{+1}\right)$ represents criss-crossing NNN couplings with strength $\delta$, and $T_{\pm1}$ denotes the translational operator across one site in the $y$ direction. Remarkably, the size of the $\bar P$ unit cell $\left(2y_R\times 2y_R\right)$ now dramatically influences the effective inter-unit cell coupling strengths and, consequently, the topological phase boundary. As shown in Fig.~\ref{fig:phasediagram}(b-d), changing the unit cell size $2y_R=2(L_y-1)$ can induce both trivial-to-nontrivial (Fig.~\ref{fig:phasediagram}(\changes{c})) and nontrivial-to-trivial (Fig.~\ref{fig:phasediagram}(\changes{d})) topological transitions.
Below, we shall describe the empirical observation of these transitions, leaving their rigorous derivations and generalizations to the next subsection.

The first scenario with topological trivial-to-nontrivial transition is demonstrated in Fig.~\ref{fig:phasediagram}(c) for illustrative parameters $t_1=0.2,t_2=0.1,\delta=0.001$. Focusing on the rightmost EB bands (circled), as $L_y$ increases from 10 [Fig.~\ref{fig:phasediagram}(\changes{c1})] to 42 [Fig.~\ref{fig:phasediagram}(\changes{c3})], the bulk EB bands (purple) close up, only to reopen at larger $L_y$ with midgap topological EB states (green) [Fig.~\ref{fig:phasediagram}(\changes{c4})].
While topological modes (green) also appear at some $L_y$ [Fig.~\ref{fig:phasediagram}(\changes{c2},\changes{c4})] in the second right band, which is not an EB band, their emergence is more erratic with a smaller gap.

Conversely, Fig.~\ref{fig:phasediagram}(d) illustrates the second scenario with a topological nontrivial-to-trivial transition for a different set of parameters $t_1=0.05,t_2=0.1,\delta=0.001$. Initially topological at small $L_y=10$ [Fig.~\ref{fig:phasediagram}(\changes{d1})], increasing $L_y$ closes up the topological gap, with the critical transition point occurring at approximately $L_y=26$ [Fig.~\ref{fig:phasediagram}(\changes{d3})], characterized by band touching. Beyond this point, the EB band transitions into a trivial topological phase, as evidenced in Fig.~\ref{fig:phasediagram}(\changes{d4}).

The possible scaling-induced ($L_y$-driven) topological transition in our setup is summarized in the topological phase diagram of Fig.~\ref{fig:phasediagram}(b), with fixed $t_2=0.1$ and variable $t_1$. The phase boundary is identified through
gap closure of the rightmost EB bands, and features an additional $L_y$-dependent curved branch, in addition to the usual $t_1=t_2$ SSH phase boundary. Intriguingly, while increasing the NN hopping ratio $t_2/t_1$ conventionally drives a trivial-to-nontrivial topological transition, for some $L_y$ (i.e., $L_y>30$ for the parameters used), the \emph{reverse} transition is observed. More interestingly, in many cases (i.e. $L_y>20$), we can obtain non-monotonic non-topological$\rightarrow$topological$\rightarrow$non-topological transitions (or vice-versa) as $t_2/t_1$ is increased. Evidently, the roles of the even/odd NN hoppings have been fundamentally modified, beyond what is possible in usual SSH-like models with further hoppings. And essentially, by tuning the system size $L_y$ in the transverse ($y$) direction, the system effectively behaves like fundamentally different models with their own topological character, without requiring \emph{any} actual modification to the NN and NNN hoppings.

\subsection{General theoretical framework for scaling-dependent EB bands and their topological transitions}
\label{sect:EBscaling}

We next rigorously establish why such scaling-dependent topological transitions occur, and how to quantitatively design them. Specifically, we show how to accurately compute the scaling dependence of generic physical EB band models far beyond the preceding examples.

\subsubsection{Construction of generic EB band models}

Generically, given any $D$-dimensional lattice model, we can promote it into a $D+1$-dimensional EB band model [Fig.~\ref{fig:barPunitcell}(a)] by replacing each physical site into a $\bar P$ unit cell, with the $\bar P$ matrix elements representing the 2-point functions of any preferred parent Hamiltonian with defective eigenspace. Mathematically, each unit cell would hence be described by $\mathbb{I}\otimes \bar P$, where $\bar P=R_\Gamma PR_\Gamma = \sum_{y,y'\in \Gamma}|y'\rangle\langle y'|P|y\rangle\langle y|$ is a EP projector $P$  restricted to a bounded physical region $\Gamma$, and $\mathbb{I}$ is the identity matrix in the unit cell of the extended space (i.e. SSH dimer for our previous example). In $1+1$ dimensions, a generic EB band model takes the form
\begin{equation}
\begin{aligned}
  H^{(\text{EBB})}_\text{Gen}&=\mathbb{I}\otimes \bar P+\sum_{x,Y,\Delta x, \Delta Y}t_{\Delta x,\Delta Y}|x+\Delta x, Y+\Delta Y\rangle\langle x,Y|\\
  &=\mathbb{I}\otimes \bar P+\sum_{k_x,Y}\sum_{\Delta x, \Delta Y}t_{\Delta x,\Delta Y}e^{i k_x\Delta x}T_{\Delta Y}|k_x,Y\rangle\langle k_x,Y|.
\end{aligned}
\label{eq:genhopping}
\end{equation}
where we have introduced the label $Y\in(1,2y_R)$ for the site index in the internal space of the $\bar P$ supercells -- $y_{R}=\lfloor Y/2\rfloor$ in our previous examples, but can be freely varied as an additional tuning knob. Here $t_{\Delta x,\Delta Y}$ is the matrix-valued amplitude of the generic hoppings across an interval $(\Delta x, \Delta Y)$, where $\Delta x$ is the number of unit cells in the extended space, and  $\Delta Y$ is the number of sites in the ``internal'' space $\Gamma$ where $\bar P$ is defined (Here, $t_{\Delta x,\Delta Y}$ is non-vanishing only for $Y,Y+\Delta Y \in \Gamma$). As emphasized in the second line, the model is intrinsically translation invariant in the extended $x$-direction (apart from possible boundaries), and different $\Delta x$ give rise to different Fourier terms $e^{ik_x\Delta x}$ in momentum space. However, translation in the $y$-direction is more nontrivial, since $\bar P$ is not a Toeplitz matrix, and we denote it explicitly as $T_{\Delta Y}$.

To summarize for the generic 1D case, starting from a translation-invariant 1D lattice with hopping amplitudes $t_{\Delta x}$, we can promote it into a 2D system with EB bands by replacing each unit cell by $\bar P$. The resultant generalized 2D EB band model is obtained by refining $t_{\Delta x}\rightarrow t_{\Delta x,\Delta Y}$ to include intra-cell hoppings $\Delta Y$, as prescribed by Eq.~\eqref{eq:genhopping}.

\subsubsection{Effective scaling-dependent 1D description of 2D EB band models}

Having described how generic 1D models can be promoted into 2D EB band models, we now derive their scaling dependence. Due to the gapped nature of EB bands, each of them behaves as an independent energy subspace.
The key realization is that projecting into any particular EB band effectively yields a non-local 1D model that is not just very different from the original 1D model, but also imbued with emergent scaling behavior.

Below, we demonstrate how to project $H^{(\text{EBB})}_\text{Gen}$ onto an EB band subspace to obtain an effective model that encapsulates its scale-dependence. This involves knowing both the left and right biorthogonal~\cite{brody2013biorthogonal} EB eigenbands, which respectively satisfy $\bar P|\phi\rangle=\bar p|\phi\rangle$ and $\bar P^\dagger|\tilde\phi\rangle=\bar p^*|\tilde \phi\rangle$. Projecting onto them, we obtain the effective 1D model
\begin{equation}
\begin{aligned}
  H^{\text{Eff}}_\text{Gen}(k_x)&=\langle \tilde \phi|H^{(\text{EBB})}_\text{Gen}(k_x)|\phi\rangle\\
  &=\bar p\,\mathbb{I}\sum_{\Delta x, \Delta Y}t_{\Delta x,\Delta Y}e^{i k_x\Delta x}\sum_Y\langle \tilde \phi| T_{\Delta Y}|Y\rangle\langle Y|\phi\rangle\\
  &=\bar p\,\mathbb{I}+\sum_{\Delta x, \Delta Y}t_{\Delta x,\Delta Y} \Omega_{\Delta Y}(L_y)e^{i k_x\Delta x}\\
  &= \bar p\,\mathbb{I}+\sum_{\Delta x}t'_{\Delta x} e^{i k_x\Delta x}.
\end{aligned}
\label{eq:genhopping2}
\end{equation}
In essence, we have reduced $H^{\text{Eff}}_\text{Gen}$ to a 1D chain with \emph{scale-dependent} effective hoppings
\begin{equation}
t'_{\Delta x}=\sum_{\Delta Y} t_{\Delta x,\Delta Y}\Omega_{\Delta Y}(L_y), \label{eq:tprime}
\end{equation}
whose $L_y$-scaling dependence is contained within the ``renormalizing" contributions
\begin{equation}
\Omega_{\Delta Y}(L_y)=\sum_Y \tilde \phi_{Y+\Delta Y}\phi_Y,\label{eq:OmegaLy}
\end{equation}
$\phi_Y = \langle Y|\phi\rangle$ and $\tilde\phi_Y = \langle Y|\tilde\phi\rangle$ being the right and left EB state amplitudes at atom $Y$.

\subsubsection{Scaling dependence in our NNN EB band model}

In the following, we apply the above formalism to showcase the precise characterization of the scaling-induced topological transition in our model $H^{(\text{EBB})}_{\text{NNN}}$ [Eq.~\eqref{eq:ham_SSH_NNN}]. The only non-vanishing physical hopping coefficients
are
\begin{equation}
\begin{aligned}
  t_{\Delta x = 1,\Delta Y=0}&=e^{i\theta}\left(
    \begin{array}{cc}
      0 & t_2 \\
      0 & 0
  \end{array}\right),\\
  t_{\Delta x = 0,\Delta Y=0}&=e^{i\theta}\left(
    \begin{array}{cc}
      0 & t_1\\
      t_1 & 0
  \end{array}\right), \\
  t_{\Delta x = -1,\Delta Y=0}&=e^{i\theta}\left(
    \begin{array}{cc}
      0 & 0\\
      t_2 & 0
  \end{array}\right), \\
  t_{\Delta x = 0,\Delta Y=\pm 1}&=e^{i\theta}\left(
    \begin{array}{cc}
      0 & \delta \\
      \delta & 0
  \end{array}\right).
\end{aligned}
\label{eq:exp2}
\end{equation}
The first three terms in Eq.~\eqref{eq:exp2} are also present in our warm-up model $H^{(\text{EBB})}_{\text{NN}}$ [Eq.~\eqref{eq:ham_SSH_NN}], and do not lead to $L_y$-scaling. This is because they all have $\Delta Y=0$, such that $\Omega_{\Delta Y=0}(L_y)=\sum_Y \tilde \phi_{Y}\phi_Y=1$ is independent of $L_y$ (see Eq.~\eqref{eq:genhopping2}).

Indeed, what makes the topological transition interestingly scaling-dependent are the $\Delta Y=\pm1$ terms, which give rise to $L_y$-dependent effective $t'_{\Delta x,\Delta Y}$ [Eq.~\eqref{eq:tprime}] hoppings. Explicitly, the non-vanishing hoppings in the effective 1D model [Eq.~\eqref{eq:genhopping2}] take the form
\begin{equation}
\begin{aligned}
  t'_{\Delta x = 1} &= e^{i\theta}
  \begin{pmatrix}
    0 & t_2 + \delta\, \Omega_{\Delta Y}(L_y) \\
    0 & 0
  \end{pmatrix}, \\
  t'_{\Delta x = 0} &= e^{i\theta}
  \begin{pmatrix}
    0 & t_1 + \delta\, \Omega_{\Delta Y}(L_y) \\
    t_1 + \delta\, \Omega_{\Delta Y}(L_y) & 0
  \end{pmatrix}, \\
  t'_{\Delta x = -1} &= e^{i\theta}
  \begin{pmatrix}
    0 & 0 \\
    t_2 + \delta\, \Omega_{\Delta Y}(L_y) & 0
  \end{pmatrix}.
\end{aligned}\label{eq:effectivet1t2}
\end{equation}
From these, we can see that a finite $L_y$ effectively ``renormalizes'' the SSH hoppings according to
\begin{equation}
\begin{aligned}
  t_1'(L_y)&=t_1 + \delta\, \Omega_{\Delta Y}(L_y),\\
  t_2'(L_y)&=t_2 + \delta\, \Omega_{\Delta Y}(L_y).
\end{aligned}
\label{t1pt2p}
\end{equation}
thereby \emph{shifting} the topological transition in an $L_y$-dependent manner. Explicitly, the phase boundary is given by $\abs{t_1'(L_y)}=\abs{t_2'(L_y)}$. This leads to a size-independent phase transition boundary $t_1=t_2$, and
\begin{equation}
t_1 + t_2 = -2\delta\Omega_{\Delta Y}(L_y),
\end{equation}
which is explicitly shifted in a manner dependent on the system size $L_y$.

\subsubsection{Scaling dependence from generic EB band hoppings}

\begin{figure*}
\centering
\includegraphics[width=0.9\textwidth]{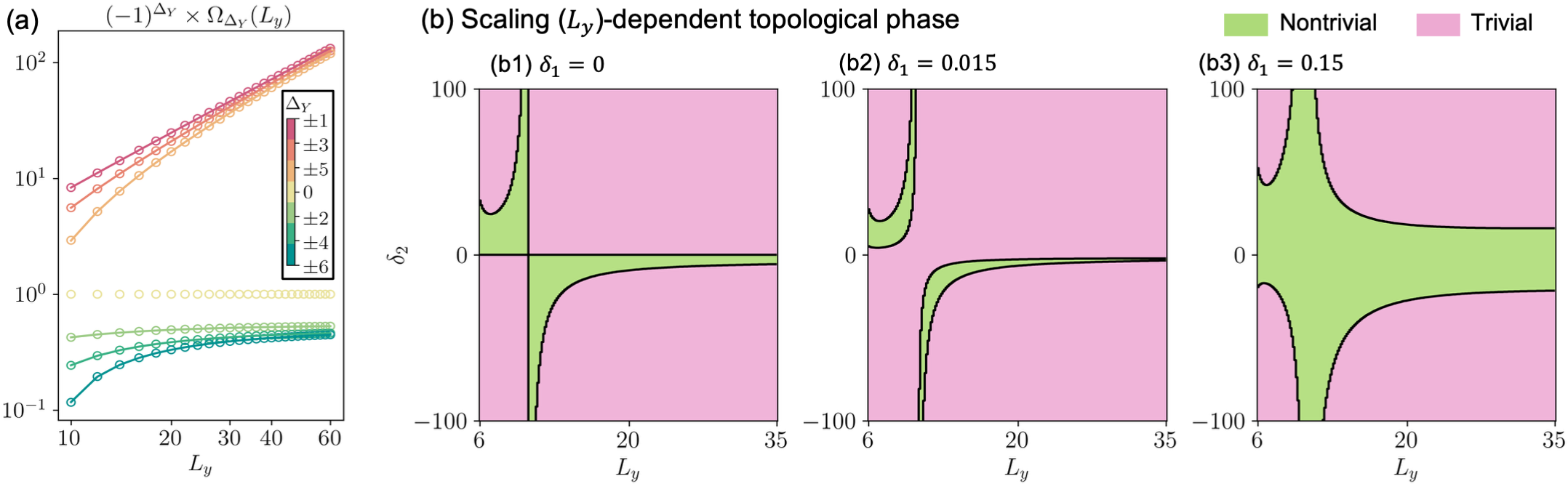}
\caption{Generalized $L_y$-scaling and diverse topological transition boundaries for extended EB band models [Eq.~\eqref{eq:ham_delta12}].
  (a) Scaling of the ``renormalizing'' factor $\Omega_{\Delta Y}(L_y)$ [Eq.~\eqref{eq:OmegaLy}], with excellent agreement between numerical results (circles) and theoretical expressions (solid curves) from Eq.~\eqref{eq:OmegaAnsatz}. Distinct $\sim L^{3/2}_y$ and $\sim L_y^{-1}+\text{const.}$ scaling behaviors are respectively observed for odd and even intra-cell hopping ranges $\Delta Y$.
(b) Qualitatively distinct topological phase diagrams in our extended model [Eq.~\eqref{eq:ham_delta12}] with effective hoppings that scale in prescribed manners [Eq.~\eqref{eq:generalized_t1t2}], for (b1) $\delta_1 = 0$, (b2) $\delta_1 = 0.015$ and (b3) $\delta_1 = 0.15$. We fix $t_1=t_2=0.1$ and $\delta_2=0.1$. The resulting topological boundaries are highly sensitive to $L_y$ and the tuning parameters $\delta_1$ and $\delta_2$, demonstrating the versatility and richness of EB band engineering beyond the simplest SSH-like lattice models. }
\label{fig:fig4}
\end{figure*}

Having pinpointed the origin of $L_y$-scaling dependence to the $\Omega_{\Delta Y}(L_y)=\sum_Y \tilde \phi_{Y+\Delta Y}\phi_Y$ term in $H^{\text{Eff}}_\text{Gen}$ [Eq.~\eqref{eq:genhopping2} and~\ref{eq:tprime}], we now derive its exact $L_y$-scaling behavior. To keep the discussion general, we shall not assume any particular form of $P(k)$ or lattice regularization, other than the requirement of geometric defectiveness i.e. that $U(k) = D(k)^{-1}$ diverges at $k\rightarrow 0$ [Eq.~\eqref{eq:P2k}].

As detailed in Methods Sect.~\ref{method:vy}, the EB basis state profiles $\phi_Y,\tilde \phi_Y$ are related to the 2-point functions as follows:
\begin{align}
\left.\phi_Y\right|_{Y=2y-1} &= u_y, \qquad \left.\phi_Y\right|_{Y=2y} = v_y \\
\left.\tilde\phi_Y\right|_{Y=2y-1} &= \mathcal{N}v_y, \quad \left.\tilde\phi_Y\right|_{Y=2y}=\mathcal{N} u_y,
\label{eq:phiUD}
\end{align}
where $u_y=(U_y - U_{y_c})/\sqrt{\displaystyle\sum_{y}(U_y - U_{y_c})^2}$ is the normalized transformed version of the 2-point function $U_y$ [Eq.~\eqref{eq:Uy}] centered at $ y_c = (1+y_R)/2$.

Given that $U_y$ scales like $U_y\sim \log L_y^2$ for $B=1$ and $U_y\sim L_y^{B-1}$ for $B>1$, and that $v_y(L_y)\;\sim\;L_y^{-B/2}$ [see Methods Sect.~\ref{method:vy}], $\Omega_{\Delta Y}\left(L_y\right)$ should also scale accordingly. Taking care of subtle normalization considerations, it can be shown that [Methods Sect.~\ref{method:OmegaDeltaY}], for the $B=2$-order parent EP used in our examples, we have the following asymptotic scaling:
\begin{equation}
\Omega_{\Delta Y}\left(L_y\right)=
\begin{cases}
  A_{\Delta Y}\, L_y^{-1}+C_{\Delta Y}\,L_y^{-2}+D_{\Delta Y}, & \text{even } \Delta Y, \\[1ex]
  E_{\Delta Y}\, L_y^{3/2}+F_{\Delta Y}\,L_y^{1/2}+G_{\Delta Y}, & \text{odd } \Delta Y,
\end{cases}\label{eq:OmegaAnsatz}
\end{equation}
which agree excellently with numerical results as presented in Fig.~\ref{fig:fig4}(a). The dependence of the coefficients is specified in Method Sect.~\ref{sec:OmegaYeven} and Sect.~\ref{sec:OmegaYeven}.
The values of the constant coefficients $A_{\Delta Y}$, $C_{\Delta Y}$, $D_{\Delta Y}$, $E_{\Delta Y}$ and $F_{\Delta Y}$, $G_{\Delta Y}$, as well as their asymptotic functional dependencies, are given in Methods Sect.~\ref{method:vy}.
Indeed, there exist clear qualitative distinctions between the nonlinear scaling behavior of $\Omega_{\Delta Y}\left(L_y\right)$ for even and odd intra-cell $\Delta Y$ hopping ranges, not just in the dominant $L_y$-scaling, but also in the subleading behavior.

\subsubsection{Design of scaling-dependent topological phase boundaries}
Finally, by exploiting the unique scaling behavior of $\Omega_{\Delta Y}(L_y)$ shown in Eq.~\eqref{eq:OmegaAnsatz} above, one can generate non-monotonic, esoteric-looking topological phase boundaries. Specifically, one constructs superpositions of $\Omega_{\Delta Y}(L_y)$ with different $\Delta Y$, such as to obtain desired asymptotic scalings with $L_y$.

As an illustration, we extend our nontrivially-scaling NNN EB band model [Eq.~\eqref{eq:ham_SSH_NNN}] to longer-ranged hoppings:
\begin{equation}
\begin{aligned}
  H^{(\text{EBB})}_{\text{extended}}(k_x, \delta_1,\delta_2) = &H^{(\text{EBB})}_{\text{NN}}\\
  &  +\delta_1\sigma_x\otimes \left(T_{-1}+T_{+1}-T_{-3}-T_{+3}\right)\\&+\delta_2\sigma_x\otimes \left(T_{-2}+T_{+2}-T_{-4}-T_{+4}\right),
\end{aligned}
\label{eq:ham_delta12}
\end{equation}
in which the EB band-induced hopping ``renormalization'' becomes
\begin{equation}
\label{eq:generalized_t1t2}
\begin{split}
  t_1'(L_y) &= t_1 + \delta_1\Bigl[\Omega_{\pm1}(L_y) - \Omega_{\pm3}(L_y)\Bigr],\\
  t_2'(L_y) &= t_2 + \delta_2\Bigl[\Omega_{\pm2}(L_y) - \Omega_{\pm4}(L_y)\Bigr].
\end{split}
\end{equation}
By containing the difference between $\Omega_{\Delta Y}$ terms with $\Delta Y$ of like parity, subleading $L_y$-scaling behavior in Eq.~\eqref{eq:OmegaAnsatz} can be accessed as
\begin{equation}
\label{eq:t1t2_scaling}
\begin{aligned}
  t_1'(L_y) &= t_1 + \delta_1 \Bigl[ (E_{\pm1} - E_{\pm3}) L_y^{3/2} \\
  &\quad + (F_{\pm1} - F_{\pm3}) L_y^{1/2} + (G_{\pm1} - G_{\pm3}) \Bigr], \\
  \\
  t_2'(L_y) &= t_2 + \delta_2 \Bigl[ (A_{\pm2} - A_{\pm4}) L_y^{-1} \\
  &\quad + (C_{\pm2} - C_{\pm4}) L_y^{-2} + (D_{\pm2} - D_{\pm4}) \Bigr].
\end{aligned}
\end{equation}
The richness of the resultant topological phase boundary is evidenced in the phase diagrams in Fig.~\ref{fig:fig4}(b1-b3), which take on qualitatively different shapes for as $\delta_1$ is slightly perturbed across $\delta_1 = 0$, $0.015$ and $0.15$.

These phase diagrams demonstrate that the band structures, particularly their topological behavior, vary non-monotonically not just with $L_y$, but also in the tunable parameters $\delta_1$ and $\delta_2$. It is worth emphasizing that these diagrams shown here represent just a small subset of the many possible $T_n$ hopping combinations in the model ansatz. Nonetheless, even this simple example already exhibits rich and interesting trends, highlighting the potential for designing a wide range of exotic phase transitions using EB band engineering, not just limited to 1D systems.

\section{Discussion}

We have proposed a fundamentally new non-Hermitian mechanism for enabling topological transitions that are controlled by the system size $L_y$. Our mechanism is based on the new paradigm of exceptional-bound (EB) band engineering, in which intra-unit cell couplings are determined by the 2-point functions of a parent exceptional point Hamiltonian. Operating independently of the well-known non-Hermitian skin effect mechanism, in which competition from skin localization leads to anomalous scaling, the transitions induced by EB band renormalization arise due to their distinct algebraic scaling properties. This distinction is crucial, as it demonstrates that non-Hermitian topology in finite-size systems is not solely dictated by skin accumulation and related concepts (i.e. the generalized Brillouin zone~\cite{yao2018edge,song2019non,yang2020non,tai2023zoology,liu2024non,li2025phase, meng2025generalized}), but can also be governed by a broader class of phenomena related to the structure of complex energy branch cuts.

More generally, EB band structure engineering and EB scaling-induced topological transitions extends to broad classes of non-Hermitian systems with different lattice geometries and hopping structures. Our approach is not limited to any particular lattice or dimensionality, being agnostic to the form of the model, as long as each unit cell can be built to contain EB degrees of freedom. We have developed a systematic approach that allow for rigorous derivation of the effective ``renormalized'' hoppings, providing a unified analytical framework for predicting scaling-dependent transitions across diverse non-Hermitian platforms, as well as potentially giving rise to new critical scaling rules and entanglement scaling paradigms.

Experimentally, EB band engineering and their scaling-induced transitions can be feasible demonstrated in any non-Hermitian physical setup that can support multiple intra-cell couplings across several or more atoms or orbitals. This include various metamaterial platforms, especially photonic crystals \changes{and electrical circuit arrays~\cite{helbig2020generalized,xiao2020non,ghatak2020observation,shang2022experimental,weidemann2020topological,lin2022simulating,wang2022non1,zhang2021acoustic,zhang2021observation,yu2021experimental,zou2024experimental,liang2022observation,zhang2024observation,xie2025non,bai2025nonlinear}}, as well as superconducting circuit-based quantum simulators~\cite{terashima2005nonunitary,su2022integrable,wang2023non,azses2024nonunitary,koh2025interacting}. In particular, the versatility of circuit setups allows for such couplings to be implemented at will.
As suggested by our topological phase diagrams, experimental smoking-guns for EB mechanisms can be formulated under diverse ranges of parameters, thanks to the sensitivity and versatility of EB band engineering.

\section{Methods}

\begin{figure}
\centering
\includegraphics[width=0.48\textwidth]{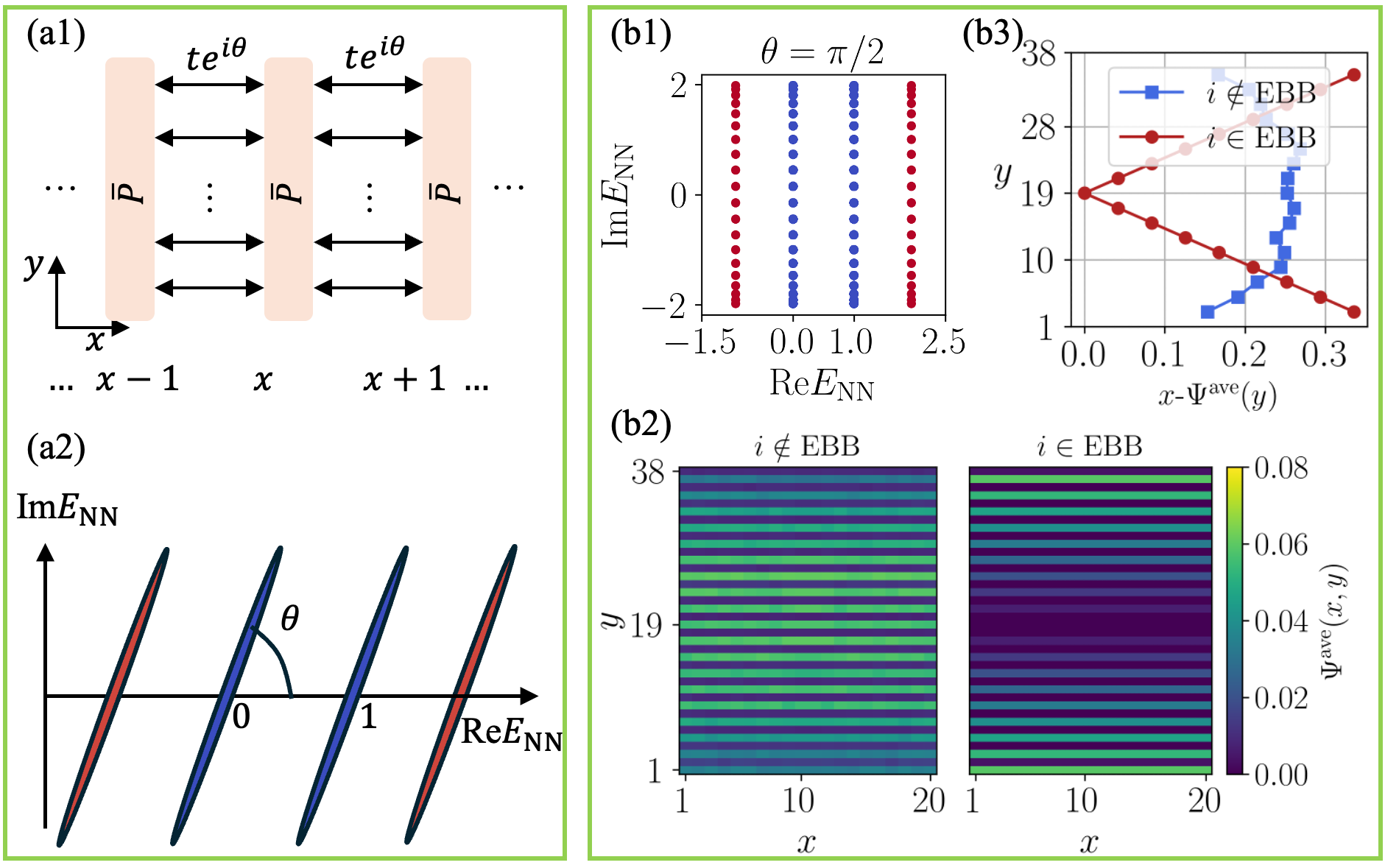}
\caption{Lattice configuration and spectral characteristics of the non-Hermitian model described by Eq.~\eqref{eq:HN_kx}. (a1) Lattice geometry comprising $L_x$ unit cells of $\bar P$ along the $y$-direction, interconnected via NN symmetric couplings $t e^{i\theta}$ along the $x$-direction. (a2) Representative energy band structure $E_{\text{NN}}$ calculated from $H_{\text{NN}}$ under open boundary conditions in the $x$-direction ($x$-OBCs).
(b1) Computed energy spectrum revealing exceptional-bound bands (EB band, red, $\chi_y=1$) and non-exceptional bulk bands (blue, $\chi_y\neq1$), distinguished by the projection fraction $\chi_y$ (Eq.~\eqref{eq:chi}). Parameters: $t=1$, $a_0=10$, $B=2$, $\theta=\pi/2$, $L_x=20$, $L_y=20$, $y_R=L_y-1$, $N=2$. (b2) Two-dimensional averaged spatial distributions $\Psi^{\text{ave}}(x,y)$ in Eq.~\eqref{eq:spatial_averaged}, shown separately for states outside (left) and within (right) the EB band. (b3) $x$-integrated spatial distributions $x\text{-}\Psi^{\text{ave}}(y)$ in Eq.~\eqref{eq:x-spatial_averaged} demonstrating distinct scaling behaviors: bulk-like distribution for non-EB bands versus linear decay with spatial localization for EB bands. Here, only the amplitudes on odd sites are plotted, as those on the even sites are almost vanishing.}
\label{fig:NNfig}
\end{figure}

\begin{figure}
\centering
\subfigure[]{\includegraphics[width=0.23\textwidth]{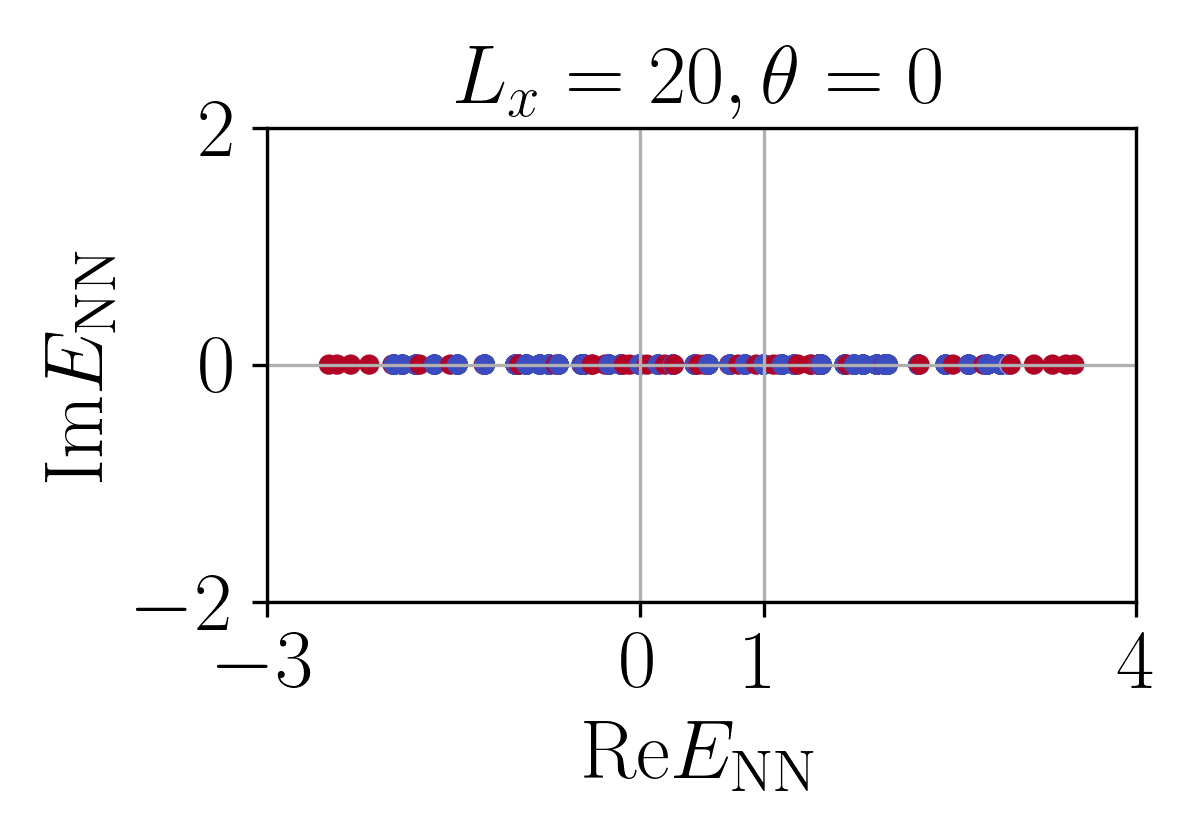}}
\subfigure[]{\includegraphics[width=0.23\textwidth]{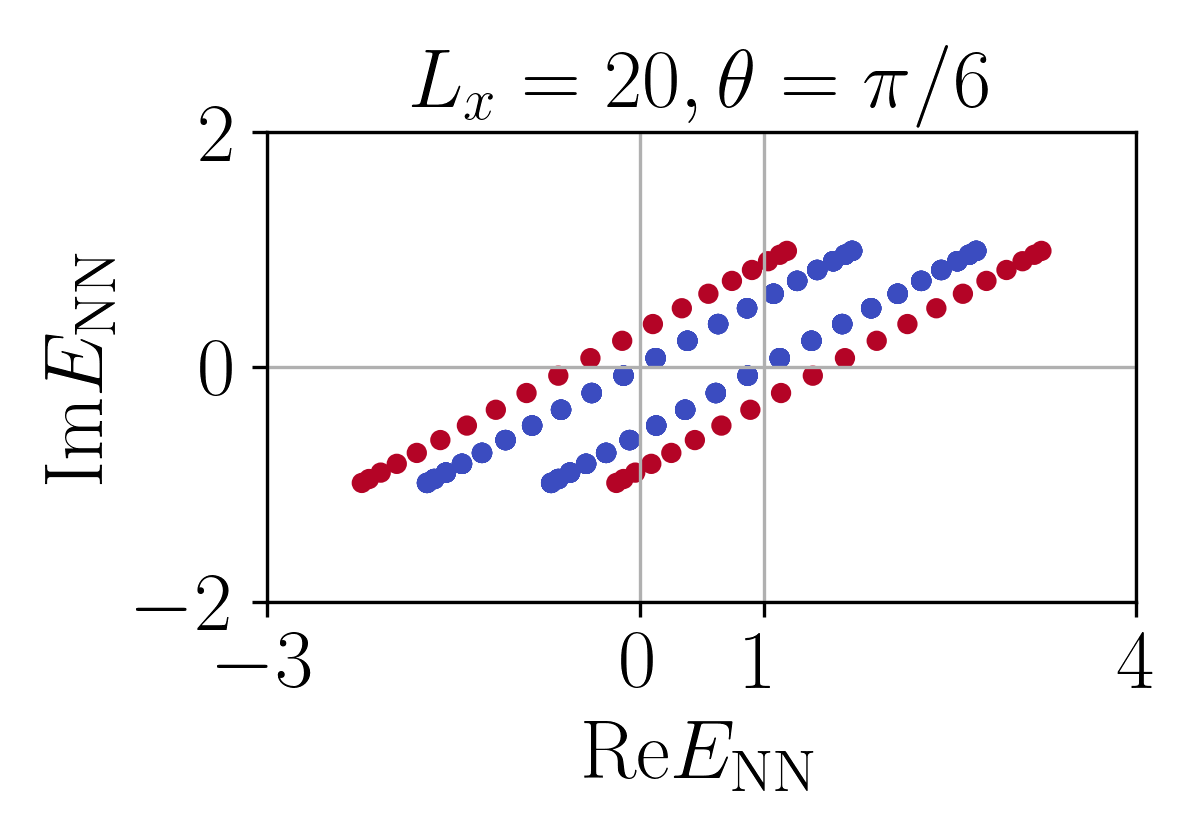}}
\subfigure[]{\includegraphics[width=0.23\textwidth]{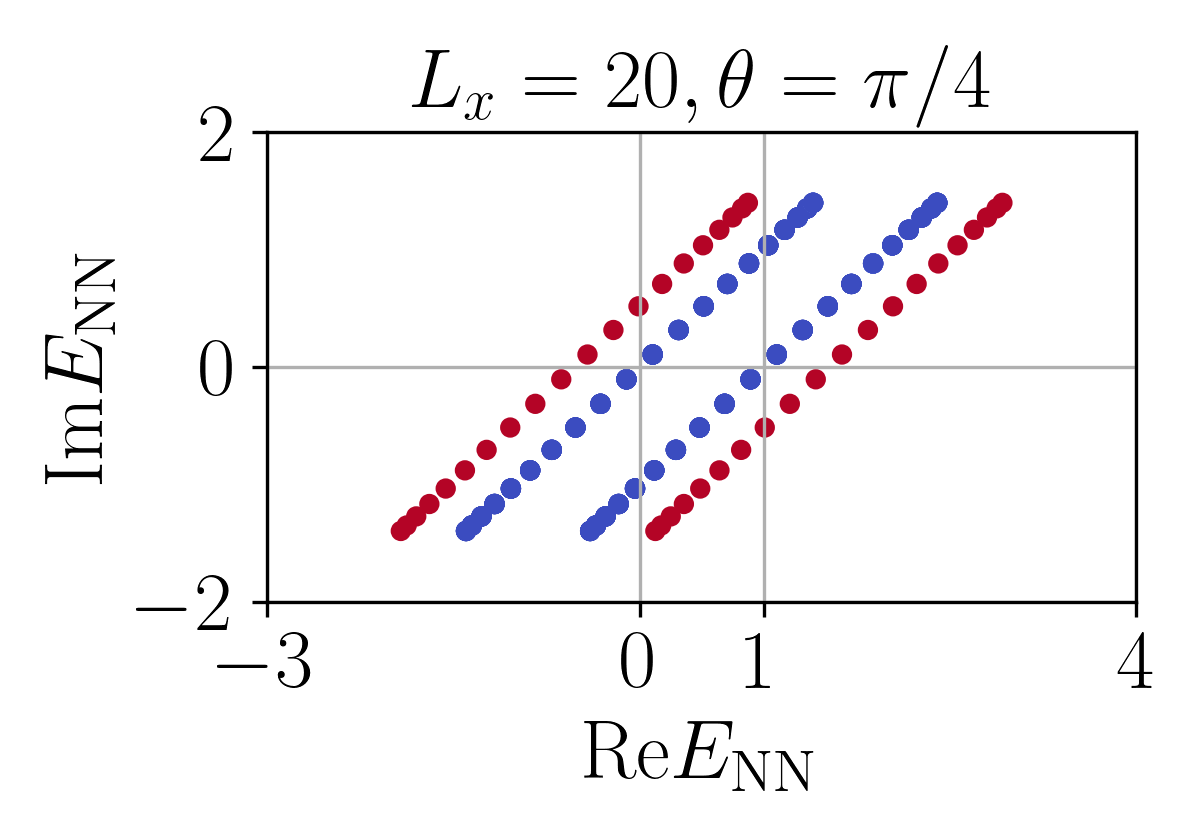}}
\subfigure[]{\includegraphics[width=0.23\textwidth]{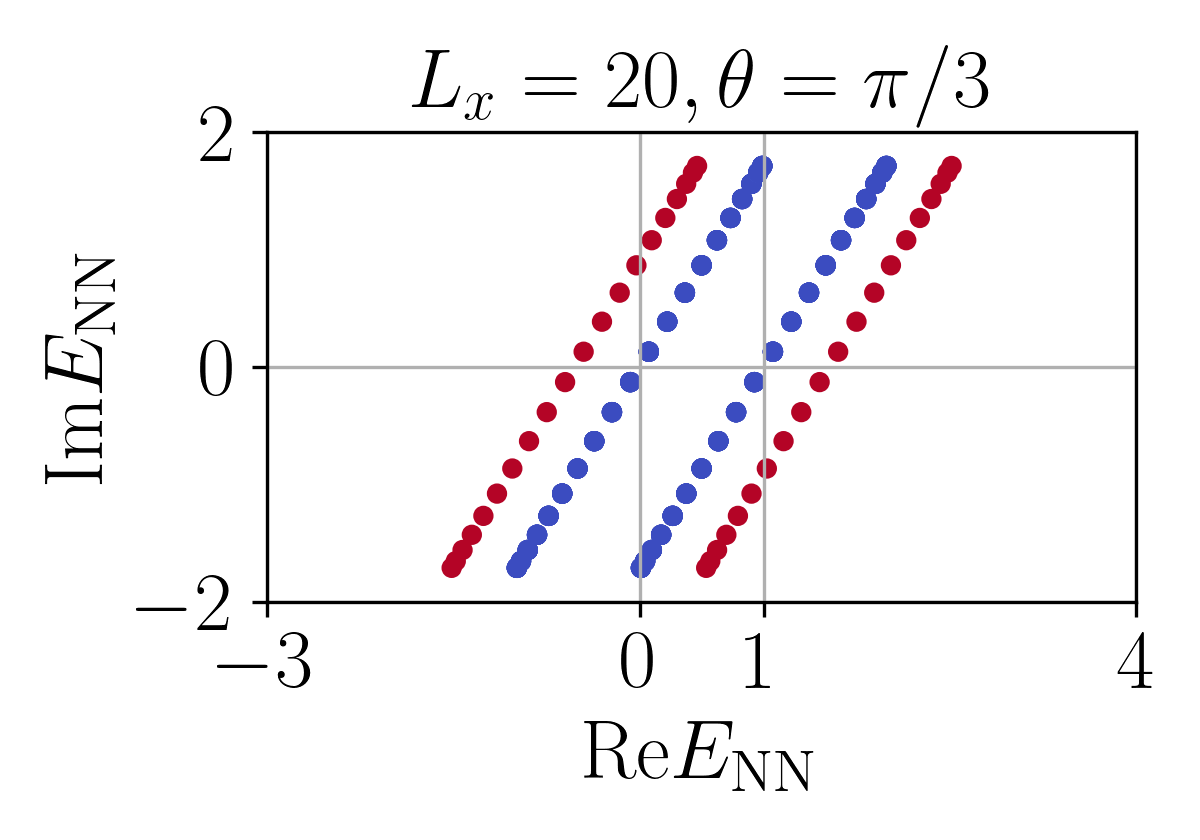}}
\caption{Rotation of the band spectra of the EB band model of Eq.~\eqref{eq:HN_kx} via $\theta$, under double OBCs. The lattice is composed of $L_y=10$ unit cells of $\bar P$ in the $y$ direction. The parameters are set as $y_R=9$, $a_0=10$ and $B=2$. The phase factor $\theta$ is varied from (a) 0 to (b) $\pi/6$, (c) $\pi/4$ and (d) $\pi/3$. The rotation of the spectrum by $\theta$ degrees in the complex plane of Re$E$-Im$E$ is evident. The color ($\chi_y$ in Eq.~\eqref{eq:chi}) is used to distinguish EB and non-EB bands.}
\label{fig:NNcouplings_phi}
\end{figure}

\subsection{Promoting EB states to EB bands -- minimal example}\label{method:EBNN}

To complement our discussion on the basic properties of an EB band lattice with intrinsic EB states, here we provide the simplest possible (single-component) example of constructing EB bands from the $\bar P$ of EB eigenstates. For simplicity, we consider the parameters $B=2$ and $y_L=1$ here. Consequently, each building block $\bar{P}$ comprises $y_R$ unit cells ($2y_R$ lattice sites), which are connected through long-range hopping interactions, depicted by the green curves previously in Fig.~\ref{fig:barPunitcell}.

With multiple $\bar P$ building blocks depicted in Fig.~\ref{fig:barPunitcell}(b) serving as fundamental building blocks arranged along the $y$ direction, we establish interconnections between these blocks via NN hopping terms $t e^{i\theta}$ along the $x$ direction, as illustrated in Fig.~\ref{fig:NNfig}(a1).
The corresponding Bloch Hamiltonian parameterized by momentum $k_x$ takes the form
\begin{equation}
H^{\text{(EBB)}}_{\text{NN}}(k_x,t,\theta) = \mathbb{I}_{2} \otimes \bar P + 2t\cos(k_x + \theta) \mathbb{I}_{2y_R}. \label{eq:HN_kx}
\end{equation}
Here $\theta$ serves as a phase rotation of the spectrum by $\theta$ degrees in the complex energy plane (Re$E$-Im$E$), as shown in Fig.~\ref{fig:NNfig}(a2) under x-OBCs. With suitable ranges of $\theta$, the resultant EB bands will not overlap, thereby precluding unnecessary subtleties in analyzing its spectral properties (see Sect. \ref{method:EBNN} later for a more detailed discussion).

The spectrum of $H_{\text{NN}}$ in Fig.~\ref{fig:NNfig}(c) consists of highly degenerate middle bands (blue) and isolated non-degenerate bands (red). To verify that they are inherited from non-EB and EB states, we compute their EB projection fraction $\chi_y$, which measures the overlap between the $x$-component of an eigenstate $\ket{\Psi(x,y)}$ where $H_{\text{NN}}|\Psi(x,y)\rangle=E_{\text{NN}}|\Psi(x,y)\rangle$ and the EB states $\ket{\phi}$ satisfying $\mathbb{I}_2\otimes \bar{P}\ket{\phi}=\bar p\ket{\phi}$. Specifically, $\chi_y$ is defined as
\begin{equation}
\chi_y=\left|\bra{\phi}\left(\sum_x\ket{\Psi_i(x,y)}^2\right)^{1/2}\right|. \label{eq:chi}
\end{equation}
The overlap criterion unambiguously classifies whether a state is in an EB band: it belongs to an EB band if and only if $\chi_y=1$, such that states with $\chi_y\neq1$ constitute the non-EB manifold.

To further confirm that EB band eigenstates inherit the spatial profiles of their EB eigenstate building blocks, we define the averaged spatial distributions of EB-band (EBB) and non EB-band (non-EBB) wavefunctions
\begin{equation}
\Psi^{\text{ave}}(x,y)=\left\{
  \begin{aligned}
    \Psi^{\text{ave}}_{\text{EBB}}(x,y)=\sqrt{\frac{\sum_{i \in\text{EBB}}\Psi^2_i(x,y)}{\sum_{i \in\text{EBB}} i}}, \\
    \Psi^{\text{ave}}_{\text{non-EBB}}(x,y)=\sqrt{\frac{\sum_{i \notin\text{EBB}}\Psi^2_i(x,y)}{\sum_{i \notin\text{EBB}} i}}.
  \end{aligned}
  \right.
  \label{eq:spatial_averaged}
\end{equation}
As shown in Fig.~\ref{fig:NNfig}(b2), while non-EB bands (left panel) exhibit typical bulk-like behavior, the EB bands (right panel) are localized in a distinctive EB-like profile, as given in Eqs.~\eqref{eq:phiUD} to \eqref{eq:Uy_B123} later. This can be seen more evidently in the $x$-integrated spatial distributions [Fig.~\ref{fig:NNfig}(b3)],
\begin{equation}
  x\text{-}\Psi^{\text{ave}}(y)=\left\{
    \begin{aligned}
      x\text{-}\Psi^{\text{ave}}_{\text{EBB}}(y)=\sqrt{\frac{\sum_{x, i \in\text{EBB}}\Psi^2_i(x,y)}{\sum_{i \in\text{EBB}} i}}, \\
      x\text{-}\Psi^{\text{ave}}_{\text{non-EBB}}(y)=\sqrt{\frac{\sum_{x, i \notin\text{EBB}}\Psi^2_i(x,y)}{\sum_{i \notin\text{EBB}} i}},
    \end{aligned}
    \right.
    \label{eq:x-spatial_averaged}
  \end{equation}
  where EB bands display linear decay and spatial localization, fundamentally different from the extended nature of non-EB bands, which is in accordance with theoretical predictions for systems with $B=2$ [Eqs.~\eqref{eq:phiUD} to \eqref{eq:Uy_B123}].

  To illustrate that the phase factor $\theta$ is to rotate the spectrum by $\theta$ degrees in the complex plane of Re$E$-Im$E$, we present the band spectrum of the lattice model in Eq.~\eqref{eq:HN_kx} with intrinsic exceptional bound states under OBCs imposed in the $x$ direction, i.e., the lattice is composed of $L$ unit cells of $\bar{P}$ in the $x$ direction. The parameters are set as $y_R = 9$, $X = 10$, $a_0 = 10$, and $B = 3$. By varying $\theta$ from 0 to $\pi$, we observe a rotation of the spectrum by $\theta$ degrees in the complex plane of Re$E$-Im$E$, as shown in Fig.~\ref{fig:NNcouplings_phi}.

  \begin{figure*}
    \centering
    \includegraphics[width=0.96\textwidth]{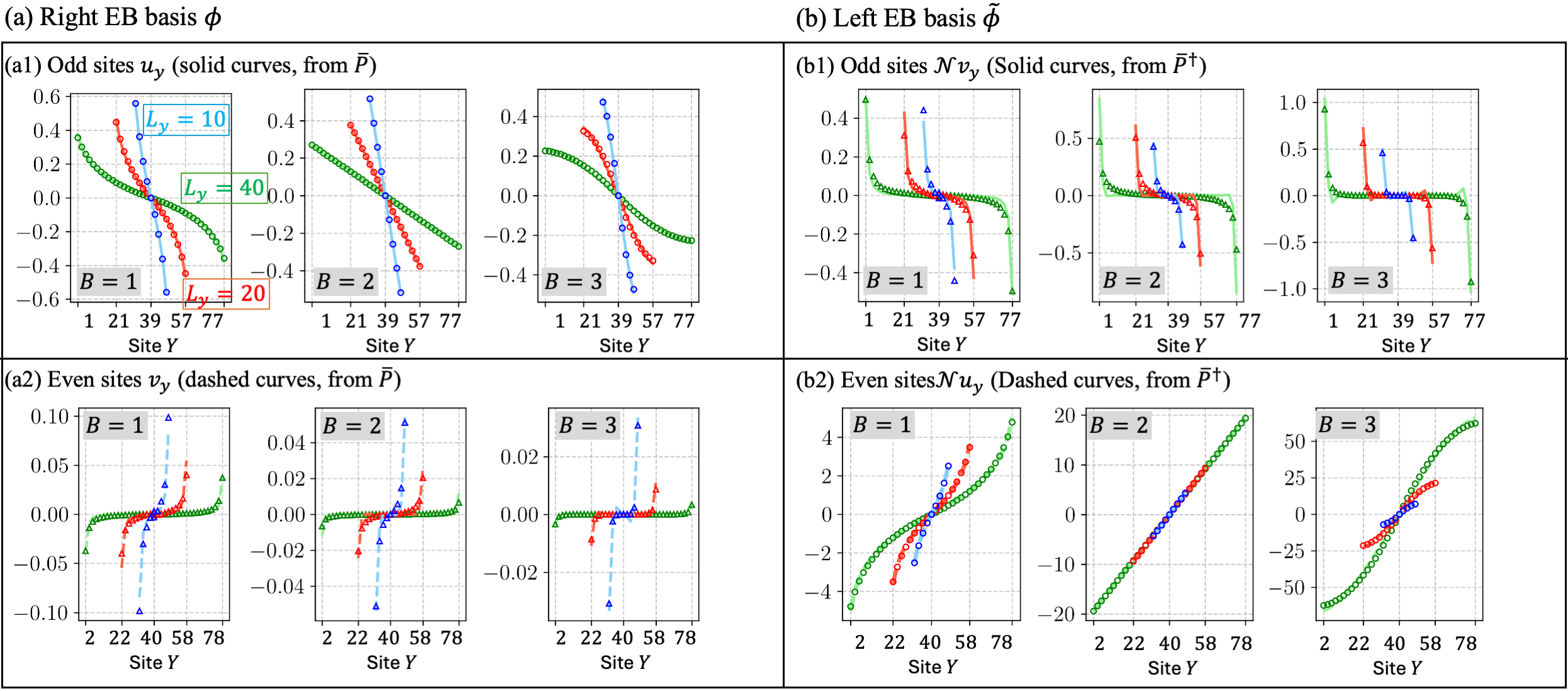}
    \caption{Scaling behavior of right and left EB basis states with respect to system size $L_y$ and site index $Y$. (a) The calculated amplitudes of right EB states show odd sites (solid curves, a1) and even sites (dashed curves, a2) closely matching $u_y$ and $v_y$ respectively, across multiple $B$ and system sizes $L_y$. (b) The left EB state amplitudes, obtained as eigenvectors of $\bar P^\dagger$, demonstrate odd sites (solid curves, b1) and even sites (dashed curves, b2) following $\mathcal{N}v_y$ and $\mathcal{N}u_y$ scaling, respectively, consistently verified for various $B$ and $L_y$ values. It is noted that $\mathcal{N}$ [Eq.~\eqref{eq:binorm}] is dependent on the system size $L_y$ and the specific $B$ value, and is serves to preserve the biorthogonal normalization $\bra{\tilde{\phi}}\phi\rangle=1$. By analyzing the scaling behaviors of these right and left EB states, we can obtain the scalings of the effective couplings $t_1'$ and $t_2'$ [Eq.~\eqref{eq:t1t2_scaling}] and predict the topological transition points. }
    \label{fig:EBbasis}
  \end{figure*}

  \subsection{Derivation of EB band scaling}
  \label{method:vy}

  \subsubsection{Scaling of EB basis states and $\Omega_{\Delta Y}(L_y)$ from correlation function behavior}
  \label{method:OmegaDeltaY}

  The EB band scaling behavior, which controls the value of $L_y$ in which topological transitions occur, originates from the scaling behavior of the biorthogonal left and right EB bases. Below we provide details on how they depend on the 2-point function $U_y$ and $D_y$ defined in Eq.~\eqref{eq:Uy}, to a high degree of accuracy.

  The 2-point functions $U_y$ and $D_y$ are characterized by the order $B$ of the parent EP singularity and the system size $L_y$~\cite{lee2022exceptional,zou2024experimental}. We illustrate their spatial scaling behavior by presenting the numerically calculated biorthogonal right and left EB states (eigenstates of $\bar P$ and $\bar P^\dagger$) with $B=1,2,3$ in Fig.~\ref{fig:EBbasis}(a) and (b), respectively. The solid (dashed) curves indicate the odd (even) sites of the calculated EB states, with different colors representing different $L_y$ values. We define two types of indices: the cell index $y$ which runs from 1 to $y_R=L_y-1$, and the site index $j$ which spans a larger range from $-y_R+1$ to $y_R$ (or equivalently, from $-(L_y-1)+1$ to $L_y-1$).

  \begin{figure}
    \centering
    \subfigure[]{\includegraphics[width=0.4\textwidth]{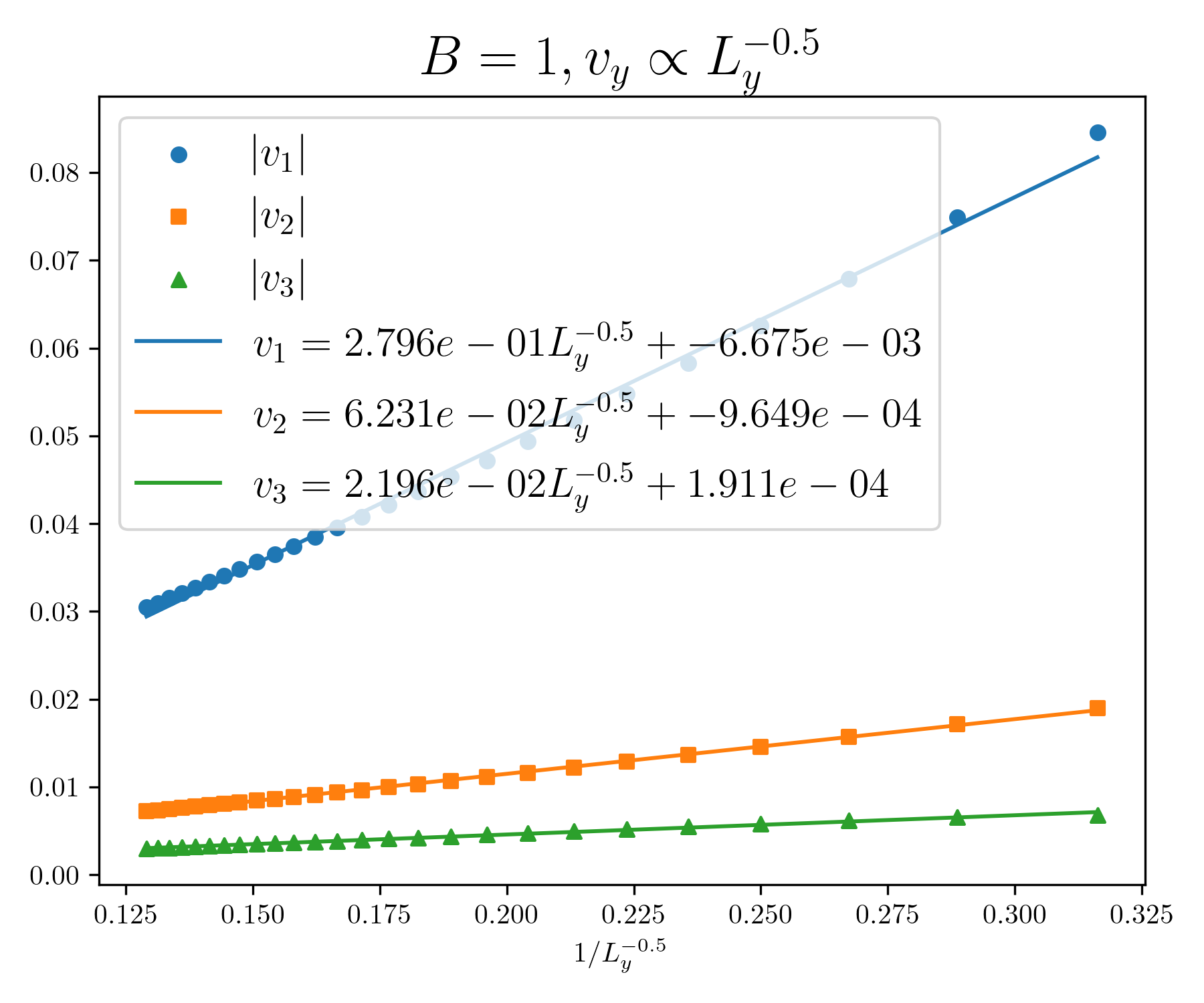}}
    \subfigure[]{\includegraphics[width=0.4\textwidth]{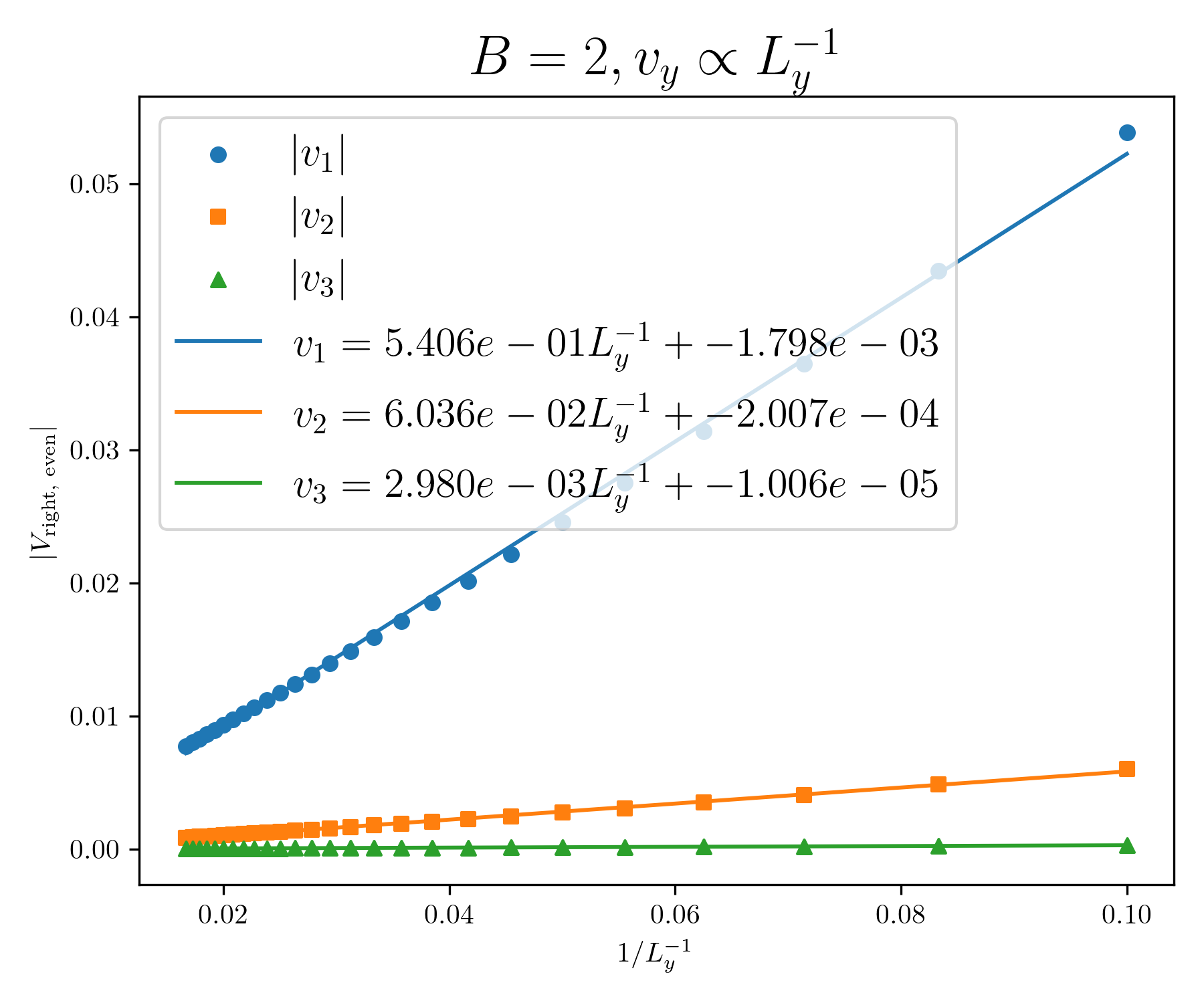}}
    \subfigure[]{\includegraphics[width=0.4\textwidth]{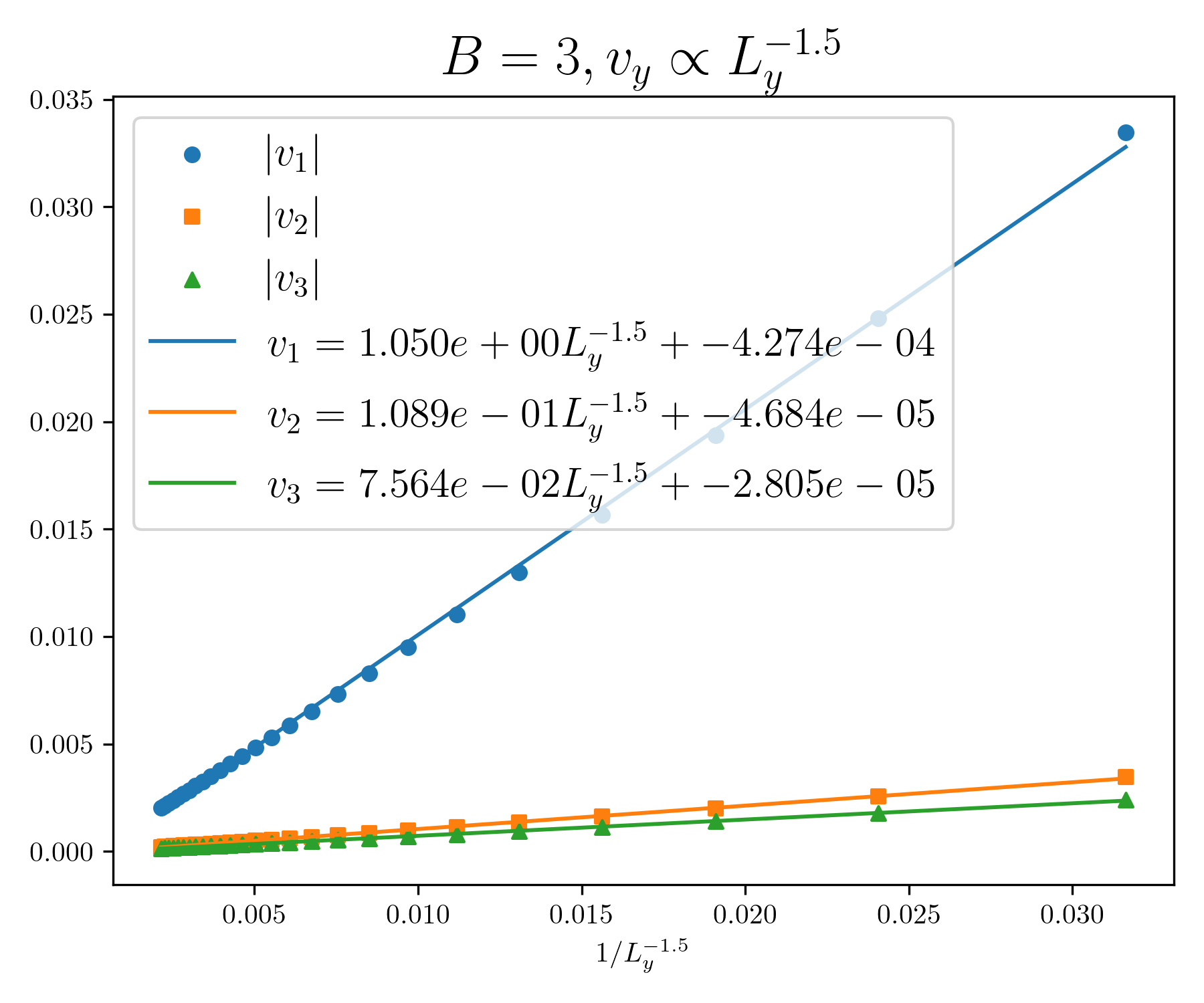}}
    \caption{The first few $v_y$ terms where $y=1,2,3$, for $B=1,2,3$. Evidently, the scaling law $v_y\sim L_y^{-B/2}$ agrees with exact numerical values (markers) excellently, and that $v_y$ decays rapidly with $y$ in all cases. }
    \label{fig:vy_scaling}
  \end{figure}

  Due to the off-diagonal form of $\bar P$, an ansatz for the EB basis states $\phi_Y,\tilde \phi_Y$ is:
  \begin{align}
    \left.\phi_Y\right|_{Y=2y-1} &= u_y, \qquad \left.\phi_Y\right|_{Y=2y} = v_y \\
    \left.\tilde\phi_Y\right|_{Y=2y-1} &= \mathcal{N}v_y, \quad \left.\tilde\phi_Y\right|_{Y=2y}=\mathcal{N} u_y,
    \label{eq:phiUD}
  \end{align}
  where
  \begin{equation}
    u_y=\frac{U_y - U_{y_c}}{\sqrt{\displaystyle\sum_{y}(U_y - U_{y_c})^2}}
    \label{eq:Uy_modified}
  \end{equation}
  is the normalized and transformed version of the 2-point function $U_y$. It vanishes at the middle site
  \begin{equation}
    y_c = \frac{1+y_R}{2}
  \end{equation}
  such as to respect the oddness of the EB basis. This ansatz exhibits excellent agreement with numerics, as shown in Fig.~\ref{fig:EBbasis}.

  Through direct integration of Eq.~\eqref{eq:Uy}, one finds that asymptotically,
  \begin{equation}
    \begin{aligned}
      U_y\big|_{B=1} &\sim 2\log \frac{L_y}{\pi y},\\
      U_y\big|_{B=2} &\sim 2\!\left(\frac{L_y}{\pi}- \pi y\right),\\
      U_y\big|_{B=3} &\sim 4\!\left(\frac{L_y}{\pi}\right)^{2}\!\left(1-\frac{\pi^{2}y^{2}}{2L_y^{2}}\right),
    \end{aligned}
    \label{eq:Uy_B123}
  \end{equation}
  with expressions for $B>3$ taking a similar form, but with a $L_y^{B-1}$ overall scaling factor. Numerically, we also find that
  \begin{equation}
    v_y(L_y)\;\sim\;L_y^{-B/2}; \label{eq:vy_numerical}
  \end{equation}
  as a function of $y$, it decays rapidly and erratically, such that only the first few terms will ever be needed.

  For the left EB states, which are eigenvectors of $H^\dagger$, their profiles can be directly related to those of the right EB states, as specified by Eq.~\eqref{eq:phiUD}. Essentially, the left ($\ket{\tilde{\phi}}$) and right ($\ket{\phi}$) EB states are spatial inversions of each other, with odd and even site amplitudes interchanged.


  The normalization factor $\mathcal{N}$ above is fixed by the biorthogonality condition $\langle \tilde{\phi} | \phi \rangle = 1$, i.e., $\langle \tilde{\phi} | \phi \rangle=\sum_Y\tilde{\phi}_{Y}\phi_{Y} =\sum_y \left(\tilde{\phi}_{2y-1}\phi_{2y-1}+\tilde{\phi}_{2y}\phi_{2y}\right)=\sum_y 2\mathcal{N}u_y  v_y=1$, which gives
  \begin{equation}
    \mathcal{N} =\left[2\sum_y u_y  v_y\right]^{-1}.\label{eq:binorm}
  \end{equation}
  \noindent Collecting the above results, the $L_y$-dependent quantity $\Omega_{\Delta Y}\left(L_y\right)$ that ``renormalizes'' the effective hoppings hence takes the form
  \begin{equation}
    \begin{aligned}
      \Omega_{\Delta Y}\left(L_y\right)&= \sum_Y \tilde \phi_{Y+\Delta Y}\phi_Y\\
      &=
      \begin{cases}
        \mathcal{N}\sum_y\left(v_{y+\frac{\Delta Y}{2}} u_y+u_{y+\frac{\Delta Y}{2}} v_y\right)&\text{even } \Delta Y, \\[0.5ex]
        \mathcal{N}\sum_y \left(u_{y+\frac{\Delta Y-1}{2}} u_y+v_{y+\frac{\Delta Y+1}{2}} v_y\right)&\text{odd } \Delta Y.
      \end{cases}
    \end{aligned}
    \label{eq:OmegaLyUyDy}
  \end{equation}

  \subsubsection{Asymptotic scaling behavior of the normalization factor \texorpdfstring{$\mathcal N$}{N}}

  For definiteness, from here on, we specialize to the $B=2$ case where the parent EP exhibits conformal symmetry (has linear dispersion). From Eqs.~\eqref{eq:Uy_B123} and \eqref{eq:Uy_modified}, we have
  \begin{equation}
    u_y\sim\frac{L_y-y}{L_y^{3/2}}, \qquad
    v_y\sim\frac{1}{L_y}+\kappa_y ,
    \label{eq:app_uv_asympt}
  \end{equation}
  where $\kappa_y$ is a constant. Keeping only the first few terms of $v_y$, we hence have
  \begin{equation}
    \begin{aligned}
      \mathcal N^{-1}
      &=2\sum_{y=1}^{L_y}u_yv_y\\
      &\simeq 
      \frac{C_0}{L_y^{3/2}}
      +\frac{C_1}{L_y^{1/2}}
      +\mathcal O\!\bigl(L_y^{-5/2}\bigr),
    \end{aligned}
    \label{eq:app_Ninv}
  \end{equation}
  with $C_0,C_1$ numerical constants that depend only on
  $\{\kappa_1,\kappa_2,\dots\}$. Taking the reciprocal of~\eqref{eq:app_Ninv} yields
  \begin{equation}
    \mathcal N
    =\alpha\,L_y^{3/2}
    +\beta\,L_y^{1/2}
    +\mathcal O\!\bigl(L_y^{-1/2}\bigr),
    \label{eq:app_N}
  \end{equation}
  where $\alpha=1/C_0$ and $\beta=-C_1/C_0^{2}$.

  \begin{figure}
    \centering
    \subfigure[]{\includegraphics[width=0.48\textwidth]{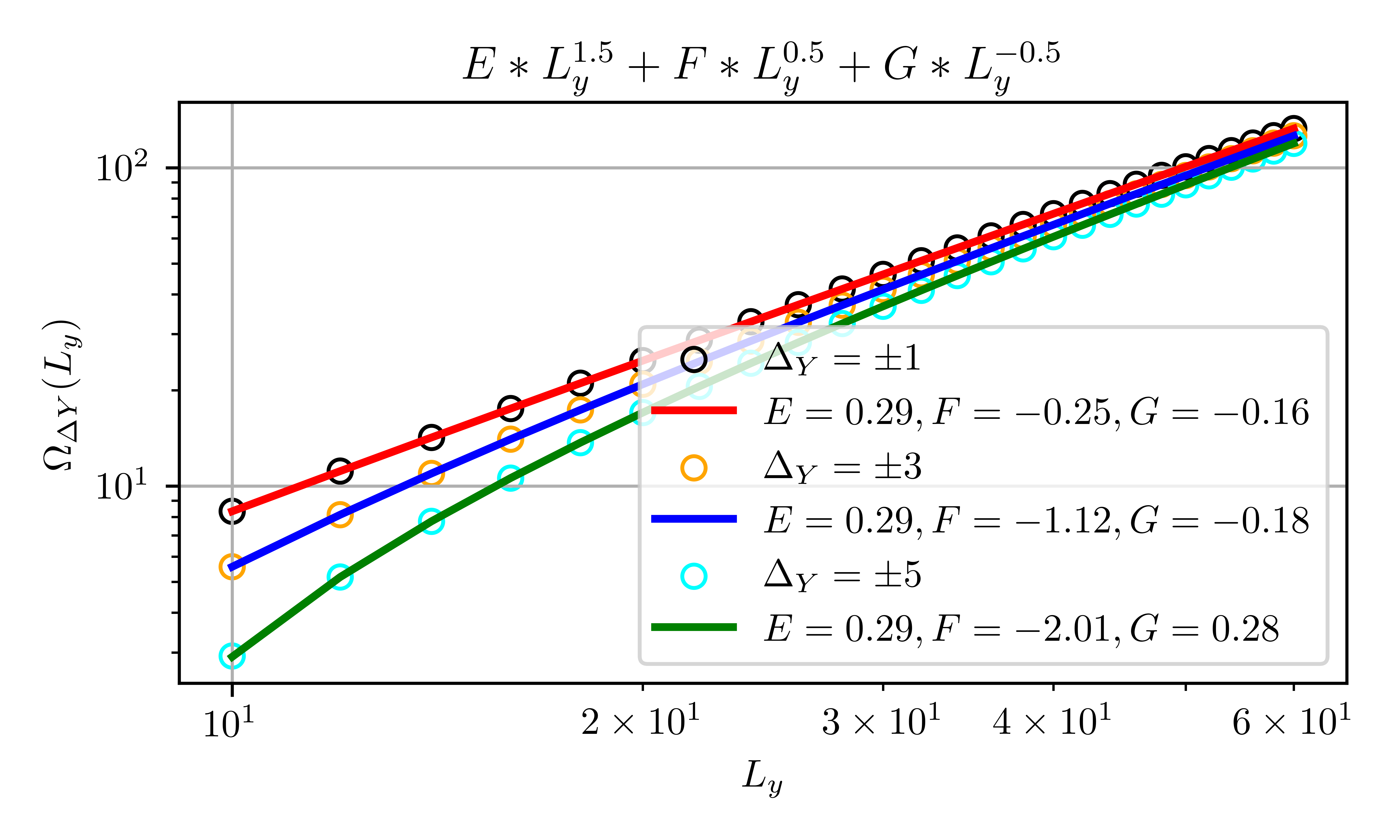}}
    \subfigure[]{\includegraphics[width=0.48\textwidth]{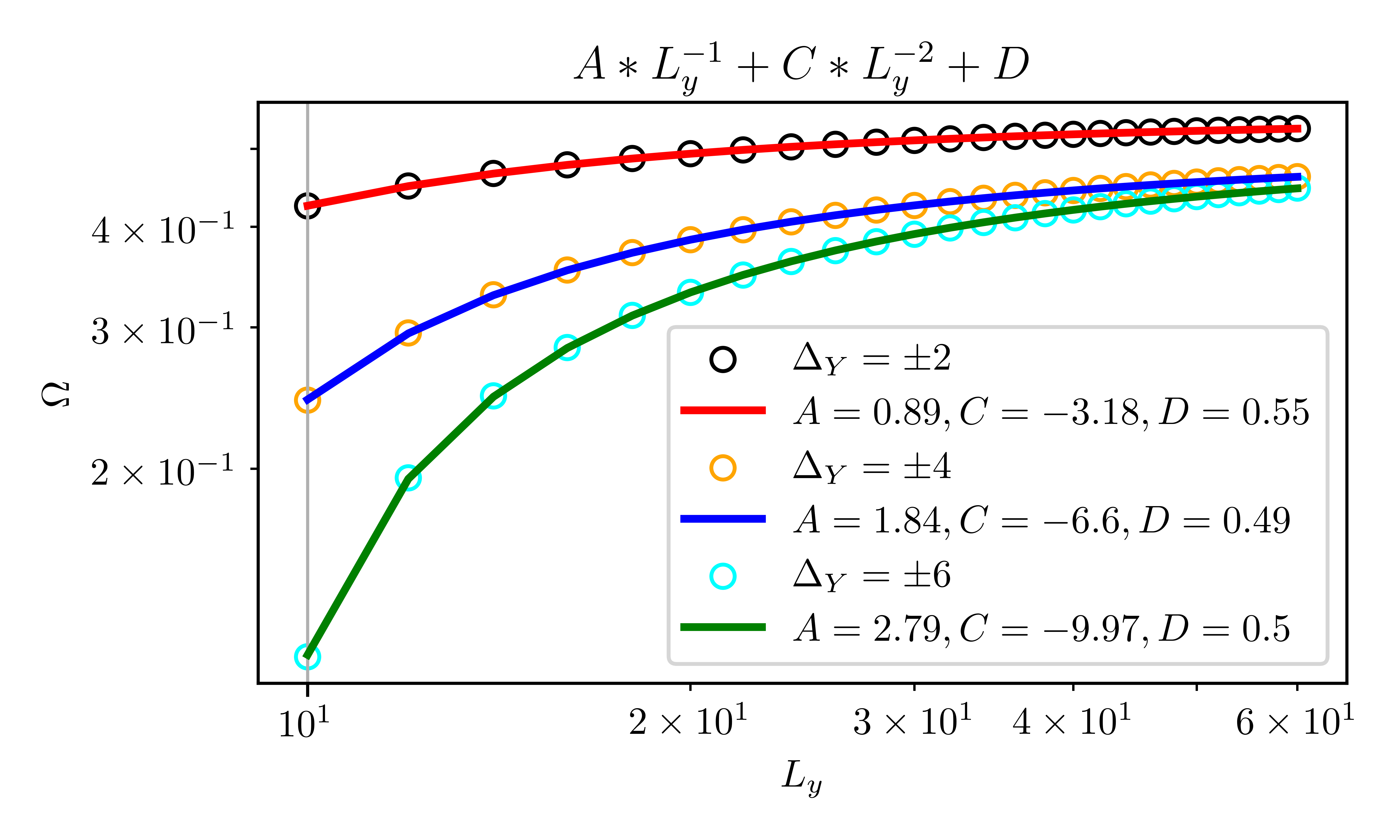}}
    \caption{The ``renormalizing'' factor $\Omega_{\Delta Y}(L_y)$ for odd (a) and even (b) displacements $\Delta Y$. Excellent fit is achieved between the numerical data (cirlces and triangles) and the ansatz Eq.~\eqref{eq:app_Omega_ansatz}.}
    \label{appfig:OmegaDeltaY}
  \end{figure}

  \subsubsection{Asymptotic scaling behavior of $\Omega_{\Delta Y}(L_y)$ with even $\Delta Y =2 s$}
  \label{sec:OmegaYeven}
  From Eq.~\eqref{eq:OmegaLyUyDy}, we have for even $\Delta Y =2 s$
  \begin{equation}
    \Omega_{2s}(L_y)
    =\mathcal N\sum_{y}
    \bigl[\,v_{y+s}u_{y}+u_{y+s}v_{y}\bigr]
    \label{eq:app_OmegaEvenDef}
  \end{equation}
  which can be approximated by its leading term in the sum:
  \begin{equation}
    \begin{aligned}
      & v_{1+s}u_1 + u_{1+s}v_1 \\[4pt]
      &\simeq\bigl[\tfrac1{L_y}+\kappa_{1+s}\bigr]
      \bigl[L_y^{-1/2}\bigr]
      +\bigl[L_y^{-1/2}\bigr]
      \bigl[\tfrac1{L_y}+\kappa_{1}\bigr] \\[2pt]
      &= \underbrace{\frac{2}{L_y^{3/2}}}_{\displaystyle\tilde C_1/L_y^{3/2}}
      \;+\;
      \underbrace{\frac{\kappa_{1}+\kappa_{1+s}}{L_y^{1/2}}}
      _{\displaystyle\tilde C_2/L_y^{1/2}}
      \;+\;\mathcal O\!\bigl(L_y^{-5/2}\bigr)\\
      &=\frac{\tilde C_1}{L_y^{3/2}}
      +\frac{\tilde C_2}{L_y^{5/2}}
      +\dots,
    \end{aligned}
  \end{equation}
  without specializing to any particular functional form for
  $\kappa_y$. Multiplying by~\eqref{eq:app_N} gives
  \begin{equation}
    \Omega_{2s}(L_y)
    =D
    +A\,L_y^{-1}
    +C\,L_y^{-2}
    +\dots,
    \label{eq:app_Omega_even}
  \end{equation}
  with coefficients $D=\alpha\tilde C_1,\quad
  A=\beta\tilde C_1+\alpha\tilde C_2,\quad
  C=\beta\tilde C_2+\alpha\tilde C_3 $.

  \subsubsection{Asymptotic scaling behavior of $\Omega_{\Delta Y}(L_y)$ with odd $\Delta Y =2s+1$}
  \label{sec:OmegaYodd}
  For odd shifts $\Delta Y =2 s+1$, Eq.~\eqref{eq:OmegaLyUyDy} gives
  \begin{equation}
    \Omega_{2s+1}(L_y)
    =\mathcal N\sum_{y}
    \bigl[u_{y+s}u_{y}+v_{y+s}v_{y}\bigr].
    \label{eq:app_OmegaOddDef}
  \end{equation}
  The summands can be approximated by the first few terms as follows:
  \begin{equation}
    \begin{aligned}
      &\frac{1}{L_y^{3}}\sum_{y=1}^{L_y-s}(L_y-y-s)(L_y-y)
      \;+\;
      \frac{1}{L_y^{2}}\sum_{y=1}^{L_y}\bigl[1+\mathcal O(L_y\kappa_y)\bigr] \\[4pt]
      &=\frac{1}{L_y^{3}}\sum_{k=s}^{L_y-1}k(k-s)
      \;+\;
      \frac{d_{1}}{L_y}
      \;+\;\mathcal O\!\bigl(L_y^{-2}\bigr) \\
      &=\frac1{L_y^{3}}
      \Bigl[\tfrac{(L_y-1)L_y(2L_y-1)}{6}
        -\tfrac{s(L_y-1)L_y}{2}
      +\tfrac{s^{2}(L_y-1)}{6}\Bigr]
      \\& \quad +\frac{1}{L_y}
      +\mathcal O\!\bigl(L_y^{-2}\bigr) \\
      &=\mathcal O(1)
      +\Bigl(1-\tfrac{s}{2}\Bigr)L_y^{-1}
      +\frac{s^{2}}{6}\,L_y^{-2}
      +\mathcal O\!\bigl(L_y^{-2}\bigr).\\
      &=\tilde D_{0}
      +\tilde D_{1}L_y^{-1}
      +\tilde D_{2}L_y^{-2}.
    \end{aligned}
  \end{equation}
  where $\tilde D_{0},D_{1},\tilde D_{2}$ can be approximated by constants. Multiplying by Eq.~\eqref{eq:app_N}, we have
  \begin{equation}
    \Omega_{2s+1}(L_y)
    =E\,L_y^{3/2}
    +F\,L_y^{1/2}
    +G\,L_y^{-1/2}
    +\dots,
    \label{eq:app_Omega_odd}
  \end{equation}
  with $
  E=\alpha\tilde D_0,\quad
  F=\beta\tilde D_0+\alpha\tilde D_1,\quad
  G=\beta\tilde D_1+\alpha\tilde D_2$.

  \subsubsection{Asymptotic scaling behavior of $\Omega_{\Delta Y}(L_y)$: Summary}

  Collecting~\eqref{eq:app_Omega_even} and~\eqref{eq:app_Omega_odd} gives the
  large-\(L_y\) behavior used in the main text:
  \begin{equation}
    \Omega_{\Delta Y}(L_y)=
    \begin{cases}
      \displaystyle
      A\,L_y^{-1}+C\,L_y^{-2}+D,
      & \text{even }\Delta Y,\\[1ex]
      \displaystyle
      E\,L_y^{3/2}+F\,L_y^{1/2}+G\,L_y^{-1/2},
      & \text{odd }\Delta Y.
    \end{cases}
    \label{eq:app_Omega_ansatz}
  \end{equation}
  These ansatz scaling expressions agree excellently with exact numerics, as verified in Fig.~\ref{appfig:OmegaDeltaY}.

  \begin{acknowledgements}
    This research is supported by the Ministry of Education, Singapore (
    MOE award number: MOE-T2EP50224-0007 (WBS no. A80035050100 ) and WBS no. A-8001542-00-00 ).
  \end{acknowledgements}

  \end{document}